\documentclass[a4paper,10pt]{article}
\usepackage{jcappub}
\usepackage{amsmath}
\usepackage{amssymb}
\usepackage[section]{placeins}
\newcommand{\dd}{\, {\rm d}}

\usepackage{cancel}
\usepackage{bigstrut}
\newcommand{\lsim}{\;\mbox{\raisebox{-0.5ex}{$\stackrel{<}{\scriptstyle{\sim}}$}
}\;}
\newcommand{\mpl}{M_{\mathrm{pl}}}
\newcommand{\gp}{{g^\prime}}
\usepackage{subfigure}
\newcommand{\gsim}{\;\mbox{\raisebox{-0.5ex}{$\stackrel{>}{\scriptstyle{\sim}}$}
}\;}
\newcommand{\pr}{^\prime}
\newcommand{\pmi}{\phi_{\rm min}}
\newcommand{\rc}{\rho_{\rm c}}

\newcommand{\dc}{\delta_{\rm c}}
\newcommand{\tdc}{\tilde{\delta_{\rm c}}}
\renewcommand\Re{\operatorname{\mathfrak{Re}}}
\newcommand{\h}{t_{\rm H}}

\newcommand{\pdp}{ {\Phi^\dagger\Phi}}

\begin{document}

\title{Dynamics of Supersymmetric Chameleons}
\author[a]{Philippe Brax}
\author[b]{Anne-Christine Davis}
\author[b]{and Jeremy Sakstein}
\affiliation[a]{Institut de Physique Theorique, CEA, IPhT, CNRS, URA 2306, F-91191Gif/Yvette Cedex, France}
\affiliation[b]{DAMTP, Centre for Mathematical Sciences, University of Cambridge, Wilberforce Road, Cambridge CB3 0WA, UK}
\emailAdd{Philippe.Brax@cea.fr}
\emailAdd{A.C.Davis@damtp.cam.ac.uk}
\emailAdd{J.A.Sakstein@damtp.cam.ac.uk}

\abstract{We investigate the cosmological dynamics of a class of supersymmetric chameleon models coupled to cold dark matter fermions. The model includes a cosmological constant in the form of a Fayet-Illiopoulos term, which emerges at late times due to the coupling of the chameleon to two charged scalars. Supergravity corrections ensure that the supersymmetric chameleons are efficiently screened in all astrophysical objects of interest, however this does not preclude the enhancement of gravity on linear cosmological scales. We solve the modified equations for the growth of cold dark matter density perturbations in closed form in the matter era. Using this, we go on to derive the modified linear power spectrum which is characterised by two scales, the horizon size at matter-radiation equality and at the redshift when the chameleon reaches the minimum of its effective potential. We analyse the deviations from the $\Lambda$CDM predictions in the linear regime. We find that there is generically a region in the model's parameter space where the model's background cosmology coincides with that of the $\Lambda$CDM model. Furthermore, we find that characteristic deviations from $\Lambda$CDM are present on the matter power spectrum providing a clear signature of supersymmetric chameleons. }
\maketitle

\section{Introduction}

The recent discovery of the acceleration of the expansion of the universe \cite{Perlmutter:1998np,Riess:1998cb} has raised some perplexing conundrums in cosmology. If the acceleration is due to a cosmological constant then why is its value so small compared to the predictions coming from particle physics? If the cosmological constant is absent, which still requires some mechanism to screen the contributions to the vacuum energy coming from the electroweak phase transition, then what is source or \textit{dark energy} driving the expansion? One of the simplest alternatives is \textit{quintessence}, a scalar field decoupled from matter slowly rolling down its potential such that its effective pressure is negative. This scenario comes with its own draw-backs: one must fine-tune the model parameters so that the acceleration begins around the current epoch and the complete decoupling from matter is unnatural. One generically needs a light scalar with mass of order $H_0$ to account for dark energy and so any coupling to matter results in a new, long ranged \textit{fifth-force} which would violate solar-system tests of gravity. This issue is not unique to quintessence, any model of dynamical dark energy introduces at least one new degree of freedom, however general relativity (GR) and a cosmological constant is the unique (up to Planck suppressed higher order curvature invariants) theory of a massless spin-2 particle \cite{Weinberg:1965rz} and so these dark energy theories are, in some sense, equivalent to modified theories of gravity. This has prompted a recent interest in the subject \cite{Clifton:2011jh}, however the problem of fifth-forces remains.

One way of avoiding this problem is to note that all of our current tests of GR have been carried out in our local neighbourhood and so there is nothing precluding large fifth-forces that are active over long ranges provided that there is some sort of \textit{screening mechanism} whereby the modifications are screened out locally. Such a mechanism was found in the form of the chameleon mechanism \cite{Khoury:2003aq,Khoury:2003rn} and since then similar mechanisms such as the symmetron mechanism \cite{Hinterbichler:2010es}, the environmentally-dependent Damour-Polyakov effect \cite{Brax:2010gi} have emerged, as has a second, independent mechanism, the Vainshtein mechanism \cite{Vainshtein:1972sx}, which is present in Galileon models \cite{Nicolis:2008in} and massive gravity \cite{Babichev:2009us}. In this work we will focus exclusively on the chameleon mechanism, which are a sub-class of scalar-tensor theories.

Chameleon theories have been well studied (see \cite{Jain:2010ka,Brax:2012yi} for some reviews), both in the context of observational signatures and their cosmological behaviour, however there has been little progress towards any sort of UV completion. Ultimately, one would like to realise these models within fundamental physics like string theory and a supersymmetric extension of these theories would be the first step towards this goal. In fact, \cite{Hinterbichler:2010wu}\footnote{see \cite{Nastase:2013ik} for a generalisation of the model and \cite{Hinterbichler:2013we} for an application to inflation.} have used the KKLT mechanism \cite{Kachru:2003aw} to find a chameleon coming from type IIB string theory compactifications at the cost of using the opposite sign in the gaugino condensation superpotential from the canonical one, which may act to decompactify the extra dimensions rather than screen fifth-forces \cite{Conlon:2010jq}. By examining supergravity breaking in a hidden sector, which can result in a direct coupling between the dark and observable sectors, the authors of \cite{Brax:2006kg,Brax:2006dc} have studied the chameleon mechanism which can emerge and have found a no-go theorem \cite{Brax:2006np}: the scale of supersymmetry breaking is so large that it renders any effects from the matter coupling negligible. This effect may be avoided if one considers only global supersymmetry where the dark and observable sectors are secluded. 

Scalar-tensor theories are IR modifications of GR and so such a bottom-up approach is sensible. In a recent letter \cite{Brax:2012mq} we have presented a general framework for embedding scalar-tensor screening mechanisms into global supersymmetry and have derived some model-independent features that arise when such theories are supersymmetrised. Within this framework, the super-chameleon couples only to cold dark matter (CDM) so that the observable and dark sectors only interact weakly via the breaking sector. This allows us to circumvent problems arising from hierarchies between the electroweak and dark energy scales. One can also introduce a matter sector, which could be in the form of the MSSM or one of its extensions. When this sector is secluded from the dark energy sector, the coupling induced by supergravity effects becomes universal and generically negligible.

With the exception of those arising from no-scale K\"{a}hler potentials, supersymmetric models with a screening mechanism are always so efficiently screened that no astrophysical signatures (such as the constraints of \cite{Hui:2009kc,Chang:2010xh,Jain:2011ji,PhysRevD.85.123006,Jain:2012tn,Mota:2012gy}) are present. Supersymmetry is always broken at finite densities, however, the seclusion of the dark and observable sectors ensures that the scale of this breaking is set by the ambient density and the model parameters and is generally well below the TeV scale associated with supersymmetric particle physics.

The large mass of supersymmetric chameleons is also associated with an equally large coupling to CDM and a low coupling to baryons. The strong Yukawa suppression of the super-chameleonic force in astrophysical situations can be circumvented at the level of linear structure formation. Indeed when large scales are involved, the large coupling to CDM particles partially compensates the Yukawa suppression of the super-chameleonic force and leads to a finite modification of Newton's constant which acts as a source for the growth rate of matter perturbations. In particular, we show that, for scales having entered the horizon before matter-radiation equality, the linear power spectrum is enhanced compared to the $\Lambda$CDM prediction in $k^2$ where $k$ is the wave-number of the large scale perturbations. On shorter scales, where non-linear effects are present and large over densities form - typically on cluster scales and below - the super-chameleon effect becomes efficient and the Yukawa suppression becomes so important that deviations from $\Lambda$CDM are expected to disappear.

In this work we will generalise the supersymmetron to an extended class of supersymmetric chameleon theories using this framework and investigate their cosmological dynamics. After introducing the general models in section \ref{sec:chamgrav} and the supersymmetric models in section \ref{sec:susycham} we will study their cosmology in section \ref{sec:cos}. These models have locally run-away scalar potentials which terminate at a supersymmetric minimum. Like all chameleons \cite{Wang:2012kj}, these models require a cosmological constant in order to account for the present day acceleration and this is particularly difficult to include in supersymmetry. Supersymmetry is broken when the vacuum energy is positive and we cannot add a cosmological constant at the level of the action. The contribution from supergravity breaking is of order $\mpl^2m_{3/2}^2$ ($m_{3/2}$ is the gravitino mass), which is far too high and must be fine-tuned away. In the letter \cite{Brax:2012mq}, we addressed this problem by introducing a mechanism by which a coupling of the super-chameleon to two scalars charged under a local $\mathrm{U}(1)$ symmetry can act to drive the mass of one of the scalars to positive values at late-times, restoring the symmetry and leaving only a Fayet-Illiopoulos (FI) term, which acts as a cosmological constant. We assume that the old cosmological constant problem in the observable and hidden sectors is resolved and fix the value of this FI term so that this cosmological constant matches the observed present-day dark energy density. Unlike scalar vacuum expectation values (VEVs), FI terms run at most logarithmically and so do not suffer from matter-loop corrections. Therefore, whilst this choice is completely arbitrary within our globally supersymmetric framework, if one can find some natural reason for this small value in a more UV complete theory then it will remain at the same order of magnitude, even at low energy scales.
In section \ref{sec:cosgen} we apply this mechanism. We indeed find a cosmological constant at late times, however at early times there are corrections to the effective potential which compete with the coupling to matter and act to negate the super-chameleon dynamics. We search the model parameter space for regions where the cosmological constant indeed appears at late time. We find that such regions are ubiquitous.
Going beyond the background cosmology level we proceed to study the linear perturbations of Cold Dark Matter. We solve the modified equation for the time-evolution of the density contrast in closed form in the matter era using modified Bessel functions. This allows us to calculate the CDM power spectrum analytically and investigate the new features due to modified gravity. In particular, when the field converges to its supersymmetric minimum sometime between matter-radiation equality and the present epoch there are three separate regimes where the spectrum exhibits different scale dependencies rather than two. We then show how the deviations from $\Lambda$CDM are characteristic in this model, with a $k$-dependent increase of power in the linear regime due to a $k$-dependent modification of Newton's constant.

\section{Chameleon Gravity}\label{sec:chamgrav}

The chameleon screening mechanism may arise from the following action,
\begin{equation}\label{eq:stmod}
 S=\int\dd^4x\sqrt{-g}\left[\mpl^2\frac{R}{2}-\frac{1}{2}\kappa^2(\phi)\nabla_\mu\phi\nabla^\mu\phi-V(\phi)+\mathcal{L}_{\rm c}(\chi_i;A^2(\phi)g_{\mu\nu})\right],
\end{equation}
which describes a scalar field coupled non-minimally to cold dark matter $\chi_i$ via the Weyl rescaled metric $\tilde{g}_{\mu\nu}=A^2(\phi)g_{\mu\nu}$; $A(\phi)$ is known as the \textit{coupling function}. In general, one may wish to couple the field to the visible matter as well and we shall discuss this later. As it stands, the action \ref{eq:stmod} describes the theory in what is known as the \textit{Einstein frame}, where the Ricci scalar is found using the \textit{Einstein frame metric} $g_{\mu\nu}$ but the non-minimal coupling results in dark matter particles following geodesics of $\tilde{g}_{\mu\nu}$, the \textit{Jordan frame metric}, so that observers in the Einstein frame infer the presence of a fifth-force
\begin{equation}\label{eq:fifthforce}
 {\bf F_{\varphi}}=\frac{\beta(\varphi)}{\mpl}{\bf \nabla}\varphi; \quad \beta(\varphi)\equiv\mpl\frac{\dd\ln A(\varphi)}{\dd \varphi},
\end{equation}
where $\varphi$ is the canonically normalised field $\dd\varphi=\kappa(\phi)\dd\phi$. The coupling function is generally taken to be of the form $A(\phi)=1+\mathcal{O}(\phi/M)+\ldots$ with $\phi\ll M$ so that perturbations in both frames do not differ too greatly. A second consequence of the non-minimal coupling is the emergence of an effective potential for $\phi$. The equations of motion are
\begin{equation}\label{eq:eom}
 \Box\varphi=\frac{\dd V_{\rm F}(\varphi)}{\dd \varphi}-\frac{\beta (\phi)}{m_{\rm Pl}}T_{},
\end{equation}
where $T$ is the trace of the energy-momentum tensor for dark matter ($T^{\mu\nu}_{}=-{2}/{\sqrt{-g}}\delta S_{\rm c}/{\delta g_{\mu\nu}}$). In fact, $T^{\mu\nu}_{}$ is not covariantly conserved in this frame since the dark matter fluid can exchange energy with the scalar; it is the Jordan frame energy-momentum tensor which is conserved $\tilde{\nabla}_\mu\tilde{T}^{\mu\nu}=0$. The trace of the energy-momentum tensor is equal to the density $\rho$ for pressureless fluids, however since this is not conserved, it is more convenient to work with the non-relativistically conserved quantity $\rc$ defined by $\rho=A(\phi)\rc$, which obeys the standard continuity equation. In the remainder of this work we shall treat $\rc$ as the conserved matter density, in which case the equation of motion \ref{eq:eom} defines an effective potential
\begin{equation}\label{eq:effpotst}
 V_{\rm eff}(\varphi)=V(\varphi)+\rc (A(\varphi)-1).
\end{equation}
The chameleon mechanism \cite{Khoury:2003aq,Khoury:2003rn} arises from potentials such as these when $V(\varphi)$ takes on a run-away form and $A(\varphi)$ is monotonically increasing such that $V_{\rm eff}(\varphi)$ has a density-dependent minimum. When large, over-dense objects such as galaxies or stars are embedded into large, low-density backgrounds (for example, the cosmological vacuum) the field will try to minimise its effective potential, the minimum of which lies at different field values inside and outside the object. Chameleons have the property that the effective mass of small oscillations about the minimum
\begin{equation}
 m_{\rm eff}(\varphi)=V_{\varphi\varphi}+\rc A_{\varphi\varphi}
\end{equation}
is an increasing function of the ambient density. In over-densities of length scale $R$ one typically has $m_{\rm eff}R\gg1$ so that the force is very short ranged and is therefore negligible. If the object is large enough that the field can reach this minimum then the force is negligible and the object is \textit{screened}, if not the mass of the field is a small perturbation about the background and the object is said to be \textit{unscreened}. For a more thorough review of chameleon screening see \cite{Khoury:2003rn,Brax:2004qh,Hui:2009kc,Jain:2012tn,PhysRevD.85.123006,Brax:2012gr}. This model has the scalar coupled to CDM particles only and so it is only necessary to screen the modifications of gravity in regions of high CDM density, laboratory \cite{Mota:2006fz,Upadhye:2012fz} and astrophysical tests \cite{Chang:2010xh,PhysRevD.85.123006,Jain:2012tn} do no apply and any field variations will not manifest in the solar-system due to a lack of any interaction with visible matter. For this reason, any screening mechanism should be able to screen in regions of CDM of density $\rc>10^6\rho_0\sim10^{-6}$eV$^4$, corresponding to the CDM density in the lightest dark matter haloes so that current cluster abundance tests \cite{Schmidt:2008tn} are satisfied\footnote{Strictly speaking, we must demand that the self-screening parameter $\chi_0$, which will be discussed below, assumes values $\lsim10^{-5}$. This would reduce to $\sim10^{-7}$ if a coupling to visible baryons were present \cite{Jain:2012tn}. We will discuss the coupling to matter induced by supergravity later and show that it is always very small.}.

\section{Supersymmetric Chameleon Gravity}

In a recent letter \cite{Brax:2012mq}, we have introduced a general framework for embedding screened modified gravity into supersymmetry and have presented a class of supersymmetric chameleons with a locally run-away potential terminating in a supersymmetric minimum at large field values. We refer the reader there for the specific details; here we shall only specialise to the class of super-chameleon models we are concerned with.

\subsection{Supersymmetric Chameleons}\label{sec:susycham}

The K\"{a}hler potential for $\Phi$ is non-canonical, which is a requirement for it to give rise to a run-away potential, whilst the dark matter fields have a canonical normalisation
\begin{equation}
 K(\Phi\Phi^\dagger)=\frac{\Lambda_1^2}{2}\left(\frac{\pdp}{\Lambda_1^2}
\right)^\gamma+\Phi_+^\dagger\Phi_++\Phi_-^\dagger\Phi_-.
\end{equation}
The self-interacting part of the superpotential is
\begin{equation}
 W=\frac{\gamma}{\sqrt{2}\alpha}\left(\frac{\Phi^\alpha}{\Lambda_0^{\alpha-3}}
\right)+\frac{1}{\sqrt{2}}\left(\frac{\Phi^\gamma}{\Lambda_2^{\gamma-3}}\right),
\end{equation}
where $\Phi=\phi+\sqrt{2}\theta\psi+\ldots$ contains a scalar $\phi$ whose modulus ultimately plays the role of the super-chameleon and $\Phi_\pm=\phi_\pm+\sqrt{2}\theta\psi_\pm+\ldots$ are chiral superfields containing dark matter fermions $\psi_\pm$. Splitting the super-chameleon field as $\phi(x)=|\phi|e^{i\theta}$ and identifying $\phi\equiv|\phi|$ from hereon, one can minimise the angular field (this is done explicitly in appendix B where a coupling to two $\mathrm{U}(1)$ charged scalars, which we will introduce later, is also examined) and define the new quantities
\begin{equation}\label{eq:scales}
 \Lambda^4\equiv\left(\frac{\Lambda_1}{\Lambda_2}\right)^{2\gamma-2}{\Lambda_2}
^4, \quad
M^{n+4}=\left(\frac{\Lambda_1}{\Lambda_0}\right)^{2\gamma-2}{\Lambda_0}^{n+4},
\quad \pmi=\left(\frac{M}{\Lambda}\right)^{\frac{4}{n}}M,\quad
n=2(\alpha-\gamma)
\end{equation}
to find the F-term potential
\begin{equation}\label{eq:Fpot}
 V_{\rm F}(\phi)=K^{\Phi\Phi^{\dagger}}\left\vert\frac{\dd W}{\dd\Phi}\right\vert^2=\left(\Lambda^2-\frac{M^{2+\frac{n}{2}}}{\phi^{\frac{n}{2}}}\right)^2=\Lambda^4\left[1-\left(\frac{\phi_{\rm min}}{\phi}\right)^\frac{n}{2}\right]^2.
\end{equation}
 The parameters $\Lambda_i$ appearing in the K\"{a}hler potential and superpotential are scales associated with non-renormalisable operators and one would expect them to be large, however, we can see that the scales governing the low-energy dynamics are $M$ and $\Lambda$. We will explore their values in detail when discussing the parameter space in section \ref{sec:cosgen}. In practice, it is easier to work with other low-energy parameters, which will be introduced later, and their relation to these parameters is given in appendix C. The index $n$ should be even and one would expect $\gamma$ and $\alpha$ to be small (but not $1$) given their presence as indices in the superpotential and so we will often consider the case $n=2$ when the need to elucidate specific calculations arises. The scale $\Lambda_3$ drops out of the dynamics when $\delta=1$, which may make fine-tuning of the dark matter mass necessary. For this reason, we will always consider $\delta\ge2$. 

When $\phi\ll\pmi$ equation \ref{eq:Fpot} reduces to the Ratra-Peebles potential
\begin{equation}
V_{\rm F}(\phi)\approx \Lambda^4 \left(\frac{\pmi}{\phi}\right)^n,
\end{equation}
which has been well studied in the context of dark energy \cite{PhysRevD.37.3406} (although one should be aware that we have not yet canonically normalised our field). At larger field values the potential has a minimum at $\phi=\pmi$ where $V(\pmi)=0$ and $\dd W/\dd \Phi=0$. Supersymmetry is therefore broken whenever $\phi\ne\pmi$.

\begin{figure}[ht]\centering
\includegraphics[width=0.5\textwidth]{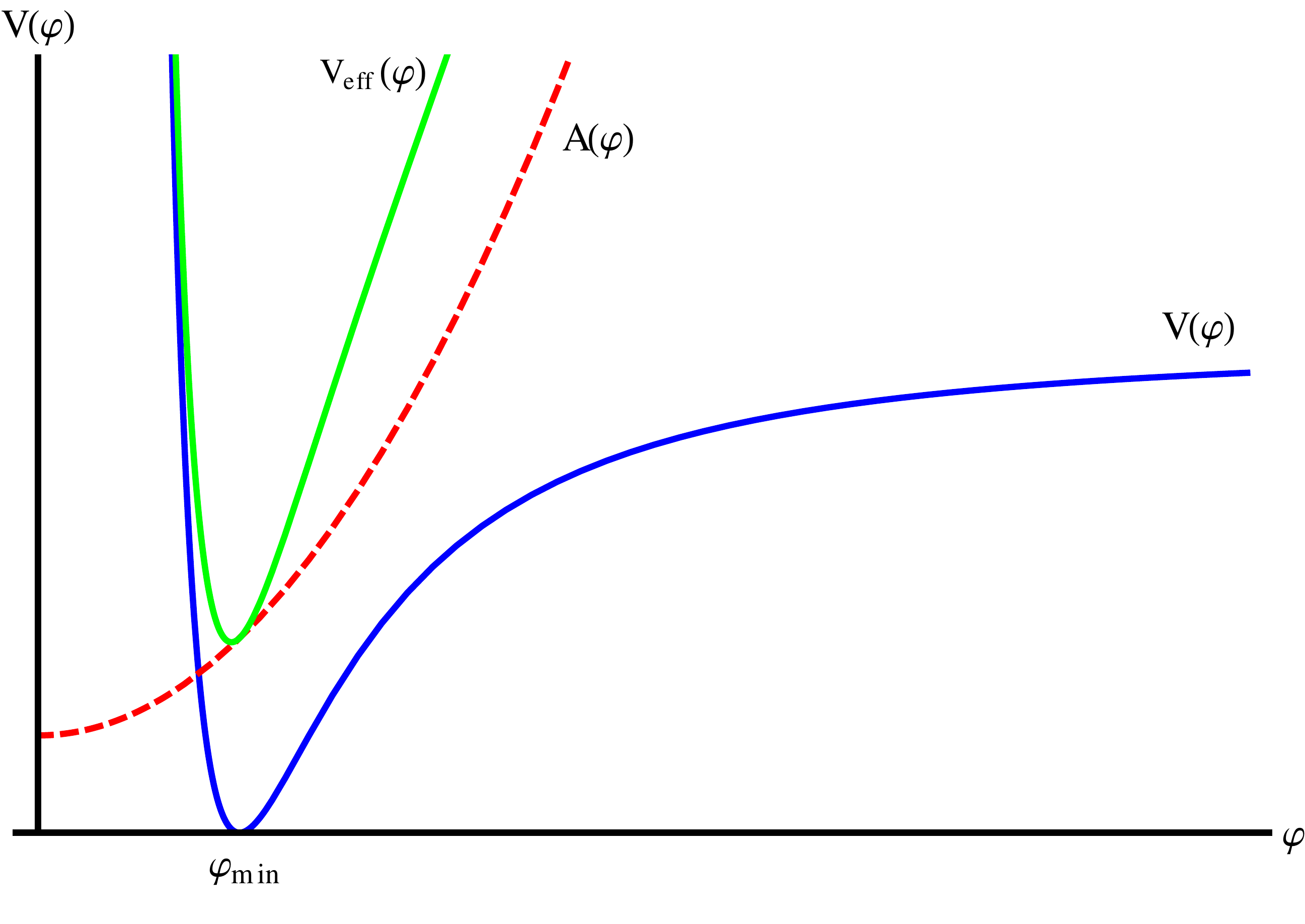}
\caption{The effective potential.}\label{fig:veff}
\end{figure}

The coupling function is found by considering the part of the superpotential containing the interactions of $\Phi$ and $\Phi_\pm$
\begin{equation}
 W_{\rm int}=m\left[1+\frac{g}{m}\frac{\Phi^\delta}{\Lambda_3^{\delta-1}}\right]\Phi_+\Phi_-,
\end{equation}
which gives a super-chameleon dependent mass to the dark matter fermions
\begin{equation}
 \mathcal{L}\supset \frac{\partial^2 W}{\partial\Phi_+\partial\Phi_-}\psi_+\psi_-.
\end{equation}
When the dark matter condenses to a finite density $\rc=m\langle\psi_+\psi_-\rangle$ this term provides a density-dependent contribution to the scalar potential resulting in the scalar-tensor effective potential $V_{\rm eff}=V+\rc (A-1)$. With the above choice of superpotential, the coupling function is
\begin{equation}\label{eq:AmodB}
 A(\phi)=1+\frac{g}{m{\Lambda_3}^{\delta-1}}\phi^{\delta}=1+\left(\frac{\phi}{\mu}\right)^{\delta};\quad \mu^\delta\equiv \frac{m\Lambda_3^{\delta-1}}{g}.
\end{equation}
As it stands, the field $\phi$ is not canonically normalised since the kinetic term in the Lagrangian reads
\begin{equation}
 \mathcal{L}_{\rm
kin}=-K_{\phi\phi^\dagger}\partial_\mu\phi\partial_\mu\phi^\dagger=-\frac{1}{2}
\gamma^2\left(\frac{|\phi|}{\Lambda_1}\right)^{2\gamma-2}
\partial_\mu\phi\partial_\mu\phi^\dagger.
\end{equation}
The normalised field is
\begin{equation}\label{eq:vphi}
\varphi=\Lambda_1\left(\frac{\phi}{\Lambda_1}\right)^\gamma
 \end{equation}
so that the coupling function \ref{eq:AmodB} becomes
\begin{equation}\label{eq:x}
A(\varphi)=1+x\left(\frac{\varphi}{\varphi_{\rm
min}}\right)^{\frac{\delta}{\gamma}};\quad x\equiv\frac{g\phi_{\rm
min}^{\delta}}{m{\Lambda_3}^{\delta-1}}
\end{equation}
and the effective potential is
\begin{equation}\label{eq:bveff}
V_{\rm eff}(\varphi)=\Lambda^4\left[1-\left(\frac{\varphi_{\rm
min}}{\varphi}\right)^{\frac{n}{2\gamma}}\right]^2+x\rc\left(\frac{\varphi}{
\varphi_{\rm min}}\right)^{\frac{\delta}{\gamma}},
\end{equation}
which is shown in figure \ref{fig:veff}.
We may then find the coupling $\beta(\varphi)$:
\begin{equation}\label{eq:betaphi}
 \beta(\varphi)=\frac{x\delta\mpl}{\gamma\varphi_{\rm
min}}\left[1+\left(\frac{\varphi}{\varphi_{\rm
min}}\right)^{\frac{\delta}{\gamma}}\right]^{-1}\left(\frac{\varphi}{\varphi_{
\rm min}}\right)^{\frac{\delta}{\gamma}-1}.
\end{equation}

The effective potential \ref{eq:bveff} is the effective low-energy potential for a scalar-tensor theory described in section \ref{sec:chamgrav} with the scalar coupled to dark matter via the coupling function $A(\varphi)$. Minimising the effective potential we have\footnote{The careful reader may notice that taking the limit $\delta=1$ gives a different equation from that found in \cite{Brax:2011bh}, which contains a typographical error.}
\begin{equation}\label{eq:phimineq}
 \left(\frac{\varphi_{\rm
min}}{\varphi}\right)^{\frac{n+\delta}{\gamma}}-\left(\frac{\varphi_{\rm
min}}{\varphi}\right)^{\frac{n+2\delta}{2\gamma}}=\frac{\rc}{\rho_\infty},
\end{equation}
where
\begin{equation}\label{eq:rhoinf}
 \rho_\infty\equiv\rc^0(1+z_\infty)^3=3\Omega_{\rm c}^0\mpl^2H_0^2(1+z_\infty)^3\equiv\frac{n\Lambda^4}{\delta x}.
\end{equation}
$z_\infty$ is an important model parameter; it is the redshift at which the field settles into its supersymmetric minimum. It turns out (as we shall see later) that this redshift controls the large scale behaviour of the modified linear CDM power spectrum, whose features change very rapidly for modes which enter the horizon after this redshift.

At zero density the field sits at the supersymmetric minimum $\varphi=\varphi_{\rm min}$ where its mass is \begin{equation}\label{eq:vphieffpot}
 m_\infty^2\equiv\frac{n\delta x \rho_\infty}{2\gamma^2\varphi_{\rm min}^2}=
\frac{3n\delta x}{2\gamma^2}\Omega_{\rm
c}^0(1+z_\infty)^3\left(\frac{\mpl}{\varphi_{\rm min}}\right)^2H_0^2.
\end{equation}
When $\rc\gsim\rho_\infty$ the field minimum is moved to smaller values by the matter coupling term and supersymmetry is broken. This is what was found in the general case studied in \cite{Brax:2012mq}. Supersymmetry is therefore broken locally in our model depending on the ambient density and $\rho_\infty$. The scale of this breaking is then set by the cold dark matter density and the model parameters, however this is generally far lower than the TeV scale associated with particle physics in the observable sector. This is one advantage of decoupling the dark and observable sectors, the dark sector does not suffer from an unnatural hierarchy of scales set by standard model particles decoupling in the visible sector. Away from the supersymmetric minimum, the field's mass is \begin{equation}\label{eq:mphi}
 m_{\rm \varphi}^2=V_{{\rm
eff},\varphi\varphi}=m_\infty^2\left[\frac{2(n+\gamma)}{n}\left(\frac{\varphi_{
\rm
min}}{\varphi}\right)^{\frac{n}{\gamma}+2}-\frac{n+2\gamma}{n}\left(\frac{
\varphi_ {\rm
min}}{\varphi}\right)^{\frac{n}{2\gamma}+2}+\frac{2(\delta-\gamma)}{n}\frac{\rc}
{ \rho_\infty}\left(\frac{\varphi_{\rm
min}}{\varphi}\right)^{2-\frac{\delta}{\gamma}}\right].
\end{equation}
Clearly $m_{\rm eff}(\varphi)>m_\infty$ when $\varphi<\varphi_{\rm min}$ and so these models are indeed chameleons with a mass at the minimum of the effective potential which depends on the matter density.

\subsection{Supergravity Corrections and Screening}

The correction to the potential coming from supergravity breaking is \cite{Brax:2012mq}
\begin{equation}
 \Delta
V_{\cancel{\mathrm{SG}}}=\frac{m_{3/2}^2\left|K_{\Phi}\right|^2}{K_{
\Phi\Phi^\dagger}}\sim\frac{m_{3/2}^2\phi^{2\gamma}}{\Lambda_1^{2\gamma-2}},
\end{equation}
which competes with the density dependent term in the effective potential (\ref{eq:bveff}). Since we focus on the branch of the potential where $\phi\le\pmi$, this term can always be neglected provided that it is far less than the density dependent term when $\phi$ has converged to its supersymmetric minimum. This requires that the supergravity corrections are negligible at densities around $\rho_\infty$ so that
\begin{equation}\label{eq:sugraphi}
 \left(\frac{\varphi_{\rm min}}{\mpl}\right)^2\ll \frac{x\rho_\infty}{\mpl^2m_{3/2}^2}.
\end{equation}
The denominator is proportional to the supergravity contribution to the vacuum energy, which is typically very large and is usually fine-tuned away whereas the numerator is proportional to the vacuum energy when the supersymmetric minimum is reached (see below), which we expect to be well below this. This condition tells us that the supersymmetric minimum must be well below the Planck scale and is simply the statement that the matter coupling and fifth-force is a low-energy, IR phenomena. It will be useful to express this condition in the alternative form
\begin{equation}\label{eq:SGB}
 \left(\frac{\varphi_{\rm min}}{\mpl}\right)^2\ll 3x\Omega_{\rm c}^0(1+z_\infty)^3\left(\frac{H_0}{m_{3/2}}\right)^2,
\end{equation}
from which it is immediately evident that $\varphi_{\rm min}\ll \mpl$ even when the gravitino mass is as low as the gauge mediated supersymmetry breaking value of $1$ eV. Using (\ref{eq:rhoinf}) equation (\ref{eq:sugraphi}) can be recast as a condition on the normalised field's mass at the supersymmetric minimum
\begin{equation}\label{eq:minfbound}
 m_\infty^2\gg \frac{3n\delta}{2\gamma^2}m_{3/2}^2.
\end{equation}
The field's mass is at least as large as the gravitino mass.

In our previous work \cite{Brax:2012mq} we have found that once supergravity corrections are accounted for, supersymmetric models such as these screen so efficiently that no astrophysical objects can be unscreened. In particular, one can show that the self-screening parameter\footnote{See \cite{Hui:2009kc,PhysRevD.85.123006,Brax:2012gr} for a more detailed discussion of the self-screening parameter.}
\begin{equation}
 \chi_0\equiv \frac{\varphi_0}{2\mpl\beta(\varphi_0)}\le \left(\frac{H_0}{m_{3/2}}\right)^2\le 10^{-33},
\end{equation}
where $m_{3/2}$ is the gravitino mass which can vary from $\mathcal{O}(\textrm{eV})$ in gauge mediated supersymmetry breaking to $\mathcal{O}(\textrm{TeV})$ in gravity mediated scenarios. When $\chi_0$ is less than the surface Newtonian potential, $\Phi_{\rm N}$, the object will be screened. However, no object in the universe has $\Phi_{\rm N}<10^{-33}$ thereby precluding any astrophysical fifth-forces.

\subsection{Coupling to baryons}

In the super-chameleon models that we are considering, the easiest way of introducing baryons or any matter particle is to introduce a secluded sector defined by its
K\"{a}hler potential $K_M$ and its superpotential $W_M$. This is a third sector on top of the dark sector and the supersymmetry breaking one. Assuming no direct interaction between the super-chameleon and matter, we have
\begin{equation}
K=K(\Phi\Phi^\dagger)+ K_{\cancel{\mathrm{SG}}}+ K_M
\end{equation}
and
\begin{equation}
W=W(\Phi)+ W_{\cancel{\mathrm{SG}}}+ W_M
\end{equation}
The mass of the canonically normalised matter fermions becomes super-chameleon dependent
\begin{equation}
m_\psi= e^{K(\Phi,\Phi^\dagger)/2M_{\rm Pl}^2}m_\psi^{(0)}
\end{equation}
where $m_\psi^{(0)}$ is the mass in the absence of super-chameleons. This leads to the coupling function in the matter sector
\begin{equation}
A_M(\varphi)=e^{\varphi^2/2M_{\rm Pl}^2}
\end{equation}
for the canonically normalised super-chameleon and the coupling to matter
\begin{equation}
\beta_M(\varphi)=\frac{\varphi}{M_{\rm Pl}}
\end{equation}
which is the coupling of a dilaton to matter. As $\phi\le \phi_{\min}$ at the minimum of the effective potential cosmologically, the largest coupling is small as long as
$\varphi_{\rm min}\ll M_{\rm Pl}$ which is guaranteed when the supergravity corrections to the scalar potential are negligible as we have seen in the previous section.

The coupling to baryons implies that the potential receives a matter dependent correction, to leading order, 
\begin{equation}
\Delta V_{\rm b}= \frac{\varphi^2}{2 M_{\rm Pl}^2} \rho_b
\end{equation}
where $\rho_b$ is the baryon density. This changes the mass of the super-chameleon by
\begin{equation}
\Delta m^2 (\varphi)= \frac{\rho_b}{M_{\rm Pl}^2}
\end{equation}
which is always negligible compared to the large mass of the super-chameleon, which we have seen must be at least as large as the gravitino mass.

Hence baryons are coupled with a very low coupling to the super-chameleon and their effect on the super-chameleon dynamics is negligible. Only CDM particles have a coupling $\beta (\varphi)={\cal O}(\frac{\varphi_{\rm min}}{M_{\rm Pl}})\gg 1$. Hence we have found
that super-chameleons are essentially only coupled to CDM particles and not to baryons. The only situations where the coupling to CDM particles is relevant are astrophysical, on the formation of large scale structure where the
Yukawa suppression of the super-chameleon force can be partially compensated by the large coupling to CDM.

\section{Cosmology}\label{sec:cos}

In this section we will examine the cosmology of these models with the aim of accounting for dark energy. We will ultimately find that a cosmological constant is required in order to match both the present day equation of state $w$ and the energy density in dark energy.

\subsection{Background Cosmology}\label{sec:backcos}

Solving (\ref{eq:phimineq}) for the minimum in the limit of both large and small dark matter density we have
\begin{equation}
 \frac{\varphi}{\varphi_{\rm min}}\approx\left\{
 \begin{array}{l l}
   \left(\frac{\rho_\infty}{\rc}\right)^{\frac{\gamma}{n+\delta}},&
\rc\gg\rho_\infty \\
    1,&\rc\ll\rho_{\infty}\\
  \end{array}\right.. \label{eq:vphimin}
\end{equation}
We can now find the contribution to the vacuum energy density
\begin{equation}
 V_{\rm eff}(\varphi)\approx\left\{
  \begin{array}{l l}
    \frac{x(\delta+n)}{n}\rc\left(\frac{\rho_\infty}{\rc}\right)^{\frac{\delta}{n+\delta}},& \rc\gg\rho_\infty \\
    x\rc,&\rc\ll\rho_{\infty}\\
  \end{array}\right. \label{eq:Veffphimin}
\end{equation}
and the mass of the field using (\ref{eq:mphi})
\begin{equation}
 \left(\frac{m_\varphi}{m_\infty}\right)^2\approx\left\{
  \begin{array}{l l}

\frac{2(\delta+n)}{n}\left(\frac{\rc}{\rho_\infty}\right)^{\frac{n+2\gamma}{
n+\delta}},& \rc\gg\rho_\infty \\
    1,&\rc\ll\rho_{\infty}\\
  \end{array}\right. .\label{eq:mphimin}
\end{equation}
Finally, one can find the equation of state for the field $w_\varphi$. In uncoupled quintessence models this is simply $P_\varphi/\rho_\varphi=-1$ when the field is at its minimum. However, in scalar-tensor theories the coupling of the field to matter results in a non-conservation of the density, and so instead one has $w_\varphi\approx -V/V_{\rm eff}$ \cite{Brax:2011qs} so that
\begin{equation}
 w_\varphi=\left\{
  \begin{array}{l l}
    -\frac{\delta}{n+\delta},& \rc\gg\rho_\infty \\
    0,&\rc\ll\rho_{\infty}\\
  \end{array}\right. .\label{eq:wmin}
\end{equation}

In order to match this with current observations we would like $w_\varphi\approx-1$ and clearly this can be achieved by taking the limits $\rc\gg\rho_\infty$, $\delta\gg n $ and imposing the condition
\begin{equation}\label{eq:viacond}
 x\delta(1+z_\infty)^3\approx3n\Omega_\Lambda^0.
\end{equation}
This corresponds to the case where $z_\infty<0$ and the supersymmetric minimum has not been reached by the current epoch. Both $n$ and $\delta$ appear as indices (or a combination of indices) in a superpotential and so we would expect them to be of similar order; taking $\delta\gg n$ is then tantamount to neglecting many lower order operators in the superpotential, making the model appear somewhat contrived. When these conditions are not met, a cosmological constant is required in order to account for the present-day dark energy observations. Unlike most models however, it is not so simple to add a cosmological constant by hand within a supersymmetric framework. Global supersymmetry is broken if $\langle V\rangle\ne0$ and so the addition of a cosmological constant to the system is non-trivial. One method is to appeal to supergravity breaking, which adds a cosmological constant of the order $\mpl^2m_{3/2}^2\gg\rho_0$ and so one must somehow fine-tune to great extent in order to arrive at the small value observed today. In this work, we shall take a different approach. If we assume that the cosmological constant problem in the matter and observable sectors is solved then we can dynamically generate a cosmological constant at late times in the form of a Fayet-Illiopoulos term provided that there exists a coupling between the chameleon and two $\mathrm{U}(1)$ charged scalars. Unfortunately, this does not remove the need for some degree of fine-tuning, since the value of the Fayet-Illiopoulos constant must be set by hand in this framework, however, this method has the advantage that this constant receives no quantum corrections from decoupling particles and so if one can find a more UV complete theory where its value is set in terms of other constants then this value would be preserved at low energy scales. The study of globally supersymmetric chameleons is aimed as a first step towards realising them within a more UV complete theory and a lot of insight can be gained by studying this mechanism.

\subsection{A Late-Time Cosmological Constant}\label{sec:cosgen}

An effective cosmological constant can be implemented by introducing two new scalars $\Pi_\pm=\pi_\pm+\ldots$ with charges $\pm q$ under a local $\mathrm{U}(1)$ gauge symmetry. These have the canonical K\"{a}hler potential
\begin{equation}
  K=\Pi_+^\dagger e^{2qX}\Pi_+ + \Pi_-^\dagger e^{-2qX}\Pi_-,
\end{equation}
where $X$ is the $\mathrm{U}(1)$ vector multiplet containing the gauge field and couple to the super-chameleon via the superpotential
\begin{equation}
 W_\pi=g^\prime\Phi\Pi_+\Pi_-.
\end{equation}
This construction gives rise to a new structure for the F-term potential as well as a D-term potential for the fields $\pi_\pm$:
\begin{equation}
V_{\rm D} =\frac{1}{2}\left(q\pi_+^2-q\pi_-^2-\xi^2\right)^2,
\end{equation}
where we have included a Fayet-Illiopoulos term $\xi^2$ which will later play the role of the cosmological constant. The new scalar potential is far more complicated with the addition of these new fields but when $\langle\pi_-\rangle=0$ it reduces to our original effective potential for the super-chameleon (\ref{eq:bveff}) plus an effective potential for $\pi_+$:
\begin{equation}\label{eq:D-term}
 V(\pi_+)=\frac{1}{2}\left(q\pi_+^2-\xi^2\right)^2+{g^{\prime}}^2\phi^2\pi_+^2;\quad \langle\pi_-\rangle=0,
\end{equation}
where in this expression we have set $\pi_+=|\pi_+|$ and will continue to do so from hereon. In appendix B we minimise the entire global F- and D-term potentials with respect to the angular fields coming from $\pi_\pm$ and show that $\langle\pi_-\rangle=0$ is indeed a stable minimum of the system.

The mass of the charged scalar $\pi_+$ (or equivalently twice the coefficient of the quadratic term in (\ref{eq:D-term})) is $m_{\pi_+}^2=\gp^2\phi^2-q^2\xi^2$. At early times the super-chameleon is small ($\ll\phi_{\rm min}$) and this mass is negative. The $U(1)$ symmetry is therefore broken ($\langle\pi_+\rangle\ne0$). However, as the cosmological field evolves towards its minimum this mass slowly increases until it reaches zero, restoring the symmetry so that $\langle\pi_+\rangle=0$. We would therefore expect $\pi_+=0$ in the late-time universe leaving us with the FI term, which plays the role of a cosmological constant. Indeed, minimising (\ref{eq:D-term}) with respect to $\pi_+$ one finds
\begin{equation}\label{eq:pi+}
q^2\pi_+^2=\left\{
  \begin{array}{l l}
    0 & \quad \phi\ge\Delta\\
    q\xi^2-{g^\prime}^2\phi^2 & \quad \phi < \Delta\\
  \end{array}\right. ,
\end{equation}
where
\begin{equation}
\Delta\equiv \sqrt{\frac{q}{{g'}^2}}\xi
\end{equation}
and $\phi=\Delta$ is equivalent to the statement $m_{\pi_+}=0$. When $\langle\pi_+\rangle=0$ equation (\ref{eq:D-term}) reduces to $V(\pi_+)=\xi^2/2$ and so we shall set $\xi\sim10^{-3}$ eV in order to match the present-day energy density in dark energy. There is no natural choice for this parameter within our globally supersymmetric framework and so this value is completely arbitrary. It is worth noting however that FI terms are largely robust to quantum corrections; when supersymmetry is unbroken they do not run and when this is not the case they run logarithmically at most \cite{Jack:1999zs}. Therefore, if one could find a natural mechanism by which a small FI term is present in a more UV complete theory, for example a suitable combination of two or more large mass scales, then its value at lower energy scales will remain at the same magnitude\footnote{Here we are assuming that the cosmological constant problem in the hidden and observable sectors is resolved.}; the same is not true of scalar VEVs, which receive large corrections from heavy particle loops.
\subsection{The Model Parameter Space}

Given the above mechanism, it is prudent to examine the model parameter space to determine the viable regions where a cosmological constant can appear Firstly, when $\langle\pi_+\rangle\ne0$ (i.e. at early times) there are corrections to the super-chameleon potential which can act to alter its cosmological dynamics. Secondly, we must ensure that the cosmological constant has the correct properties to reproduce current observations. We require the cosmological constant to appear before the present epoch and a necessary condition for this is
\begin{equation}\label{eq:mincond}
 \phi_{\rm min}>\Delta\quad\textrm{or equivalently}\quad \left(\frac{M}{\Lambda}\right)^{\frac{4}{n}}>\left(\frac{q}{2{g^\prime}^2}\right)^{\frac{1}{2}}\frac{\xi}{M}.
\end{equation}
This is an additional constraint that must be imposed on the model parameters. Furthermore, if our model is to produce the correct imprint on the CMB then the cosmological constant must be generated before last scattering. We shall do this by imposing that the cosmological density $\rho_\Delta$ (given in (\ref{eq:rhoi})) at which the $\mathrm{U}(1)$ symmetry is restored is greater than $1$ eV$^4$.

\subsubsection{Corrections to the Scalar Potential}
At late times (defined by the time at which $\phi=\Delta$) we have a FI cosmological constant, but at earlier times the non-zero VEV of $\pi_+$ induces corrections to the effective potential for $\phi$:
\begin{equation}\label{eq:dcorr}
 V_{\rm corr}=\frac{{g^\prime}^2\xi^2}{q}\phi^2-\frac{{g^\prime}^4}{2q^2}\phi^4.
\end{equation}
These corrections compete with the density-dependent term coming from the dark matter coupling and therefore act to negate the chameleon mechanism. When they are important, they lead to a new, density-independent minimum and since the magnitude of the density dependent term decreases as the dark matter redshifts away it is possible to have a scenario where the field gets stuck at the new minimum and the cosmological constant is never generated. At first glance, one may be concerned that the correction proportional to $-\phi^4$ results in a potential that is unbounded from below, but this form of the potential is deceptive. If one were to consider allowing the field to run away down this potential then at some point we would be in a situation where $\phi>\Delta$ and these corrections are no longer present; what looks like an unbounded potential is in fact a hill in the global potential.

There are several possible scenarios involving these corrections, which either allow or preclude the generation of a cosmological constant depending on the model parameters. If the corrections are negligible compared to the density dependent term throughout the entire time that $\phi<\Delta$ then they are never important to the model dynamics and vanish once $\phi>\Delta$. If, on the other hand, the corrections are important before they vanish then their dynamics must be included. However, if $\phi$ can still pass $\Delta$ then a cosmological constant can still be generated since the corrections vanish after $\Delta$ is passed. If the only important correction is the quadratic one then a minimum always develops and therefore the cosmological constant will only be generated if the field value at this minimum is larger than $\Delta$. If either the quartic correction or both corrections simultaneously are important then the potential may or may not develop a minimum. If no minimum develops then the field will eventually pass $\Delta$ since the potential takes on a (locally) run-away form. If a minimum does develop then we again require the field value at this minimum be larger than $\Delta$ in order to generate a viable cosmology. The exact details of how one can determine which scenario is applicable to a certain choice of parameters and whether or not the dynamics are affected to the extent that the model is not viable are given in full detail in appendix C. Below we shall only present the resulting parameter space once every possible scenario is taken into account.

\subsubsection{Low-Energy Parameters}

In order to classify the parameter space into viable regions we will need to
derive certain conditions on combinations of the model parameters and so it is
important to know which parameters are fixed in terms of certain combinations of
the others. It will be sufficient to examine the position of the minima and the
values of $\phi$ relative to $\Delta$ and at no point will we need to use the
dynamics of $\varphi$. For this reason, we will work exclusively with the field
$\phi$ and not its canonically normalised counterpart since this avoids
unnecessary powers of $\gamma$. We have already seen in section
\ref{sec:susycham} that three of the underlying parameters $\Lambda_i$
($i=0,1,2$) combine to form two derived parameters $M$ and $\Lambda$. What are
observable however are the low-energy parameters
${n,\delta,\gamma,x,\mu,z_\infty,\gp}$, which are either combinations of $M$ and
$\Lambda$ or indices that appear in the low-energy effective potentials
(\ref{eq:AmodB}) and (\ref{eq:D-term}); $\mu$ is a combination of the underlying
parameter $\Lambda_3$ and the dark matter mass $m$. It will prove useful to
introduce the parameter
\begin{equation}\label{eq:Gdef}
 G\equiv \gp/\sqrt{q}.
\end{equation}
A static analysis therefore probes the six dimensional parameter space
${n,\delta,x,\mu,z_\infty,G}$ and leaves $\gamma$ unspecified.

In what follows, we will be interested in regions of parameter space where the background cosmology is viable and the parameters themselves assume \textit{sensible} values. In order to decide exactly what is meant by ``sensible'' it is instructive to pause and think about their physical significance. $n$ and $\delta$ are indices (or are combinations of indices) that appear in a superpotential and so these should naturally have values close to $1$ as argued in section \ref{sec:backcos}. $g\pr=\sqrt{q}G$ is a $\mathrm{U}(1)$ coupling constant that appears in the coupling of the charged fields to $\phi$ and so we would expect values of $\mathcal{O}(10^{-2}-10^{-3})$ so that the theory is not strongly coupled. Values much smaller than this would be tantamount to fine-tuning. Similarly, $x$ parametrises the ratio of the vacuum energy density to the matter density when the field has converged to its supersymmetric minimum. The energy density due to the field today is (see equation (\ref{eq:Veffphimin})) $V_{\rm eff}(\pmi)=x\rc$, which must be less than $\xi^4$ so that the dominant contribution to dark energy comes from the cosmological constant and so we require $x\lsim\mathcal{O}(1)$. Na\"{i}vely, one might argue that $x$ should be small since it also parametrises the coupling to matter (see equations (\ref{eq:x}) and (\ref{eq:betaphi})) and so directly controls the enhanced gravitational force. However, we have already seen that supergravity corrections ensure that all astrophysical fifth-force effects are screened and so this argument does not apply.

Finally, we are left $\mu=m^{\frac{1}{\delta}}\Lambda_3^{\frac{\delta-1}{\delta}}$. When $\delta=1$ this is simply the dark matter mass and thus varying $\mu$ is tantamount to fine-tuning the dark matter mass so that we get an acceptable cosmology. When $\delta\ne 1$ however we are free to fix the dark matter mass and what we are really varying is $\Lambda_3$. In this sense we are not fine-tuning when we vary $\mu$ but are in fact scanning the space of viable cosmologies as a function of $\Lambda_3$. For this reason, we shall always fix $\delta\ne1$. Now the dark matter mass can be any where from $\mathcal{O}(\textrm{eV})$ to $\mathcal{O}(\textrm{TeV})$ depending on the model and $\Lambda_3$ appears as a mass scale in the underlying supersymmetric theory and so we would naturally expect it to be large (at least compared to the scales involved in the low-energy dynamics). Hence, in what follows we will treat anywhere in the region $\mathcal{O}(\textrm{eV})\lsim\mu\lsim\mathcal{O}(\mpl)$ as sensible.

\subsection{Constraints on the Parameter Space}

Given the above considerations and the procedure for dealing with corrections to the effective potential in appendix C we are now in a position to explore the parameter space at the background level.

We have performed a thorough investigation into the exact effects of varying each of the six parameters on the cosmology and can find a large region where the parameters are indeed sensible and the background cosmology is viable. It is difficult to gain any insight from the equations since they are all heavily interdependent in a complicated fashion and a large number of plots can be misleading since they can change very abruptly when a single parameter is varied by a small amount. For these reasons, here we shall simply describe the effect of varying some of the more constrained and less interesting parameters and present only a few two-dimensional cross-sections once these have been fixed at sensible values.

\begin{figure}[ht]\centering
 \includegraphics[width=0.75\textwidth]{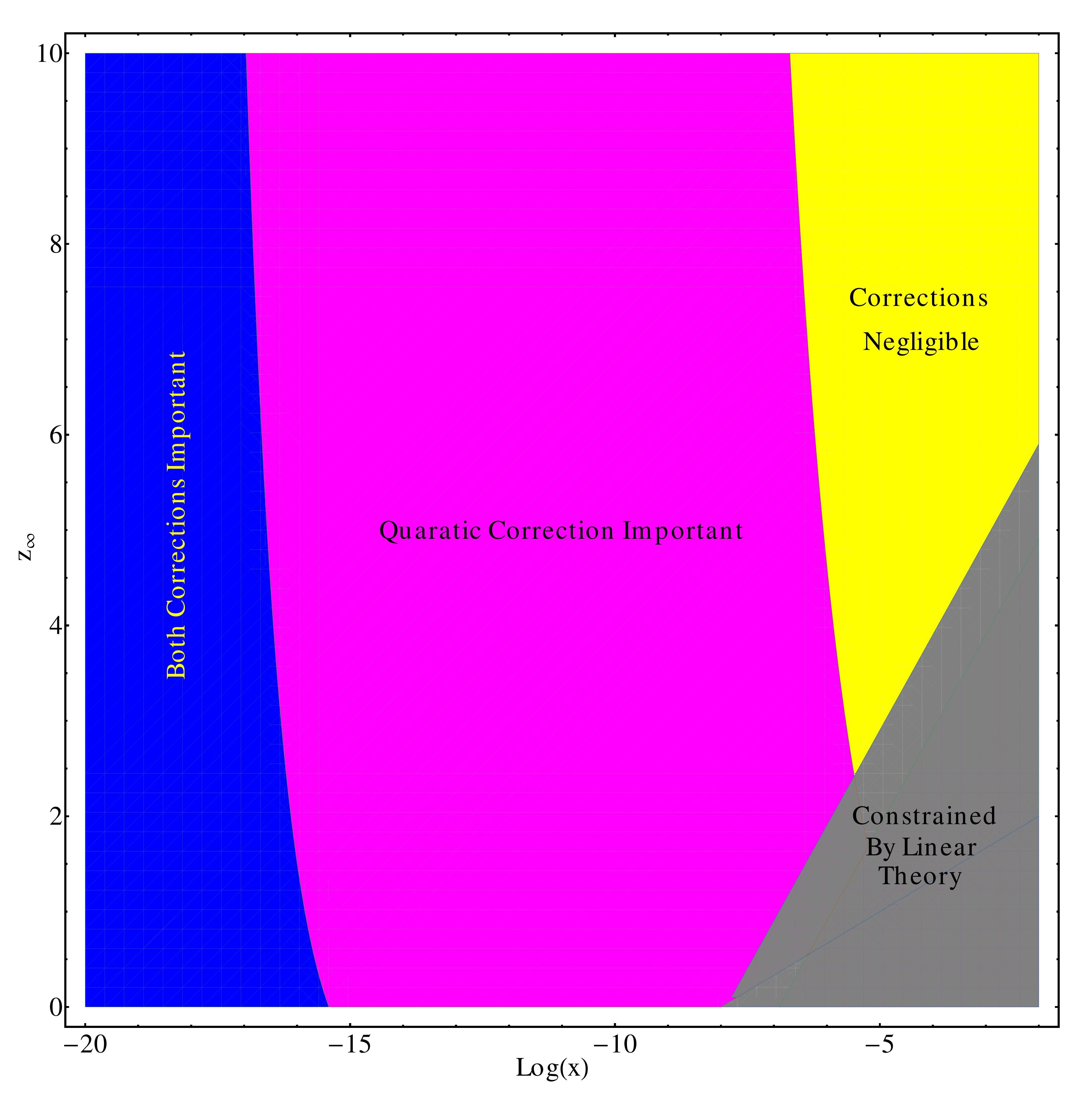}
\caption{The various regions in the $z_\infty$-$\log(x)$ plane with $n=\delta=2$, $\mu=10^3$ TeV and $G=10^{-2}$. The yellow region shows the parameter range where the corrections are negligible. The magenta region shows the ranges where the quadratic correction is important, the dark blue region where both corrections are important. The grey region corresponds to parameters where the model deviates from $\Lambda$CDM at the level of linear perturbations.}\label{fig:zx}
\end{figure}

Let us begin with the indices. $n$\footnote{Note that $n=2(\gamma-\alpha)$ and
so only even values of $n$ are allowed.} and $\delta$ have very similar effects:
if their value is increased whilst fixing the other five parameters then the
region of parameter space where the corrections can be neglected will increase.
Being indices, these should not stray too far from $\mathcal{O}(1)$ and so their
effects are far less pronounced than the other parameters, some of which may
vary over many orders of magnitude. Hence, from hereon we will fix $n=\delta=2$.
In figure \ref{fig:zx} we plot
the $z_\infty$-$\log(x)$ plane with $\mu=10^3$ TeV and $G=10^{-2}$ corresponding
to what we have argued above are sensible values. For the sake of brevity we
will set $z_\infty=5$ from hereon. This choice is completely arbitrary and
different choices may give rise to very different cross-sections of parameter
space, however, the region where the corrections are negligible is both
ubiquitous and generically large when $z_\infty\gsim0$ and so it is not
necessary to scan this parameter in great detail in order to narrow down a
viable region.

\begin{figure}[ht]\centering
 \includegraphics[width=0.75\textwidth]{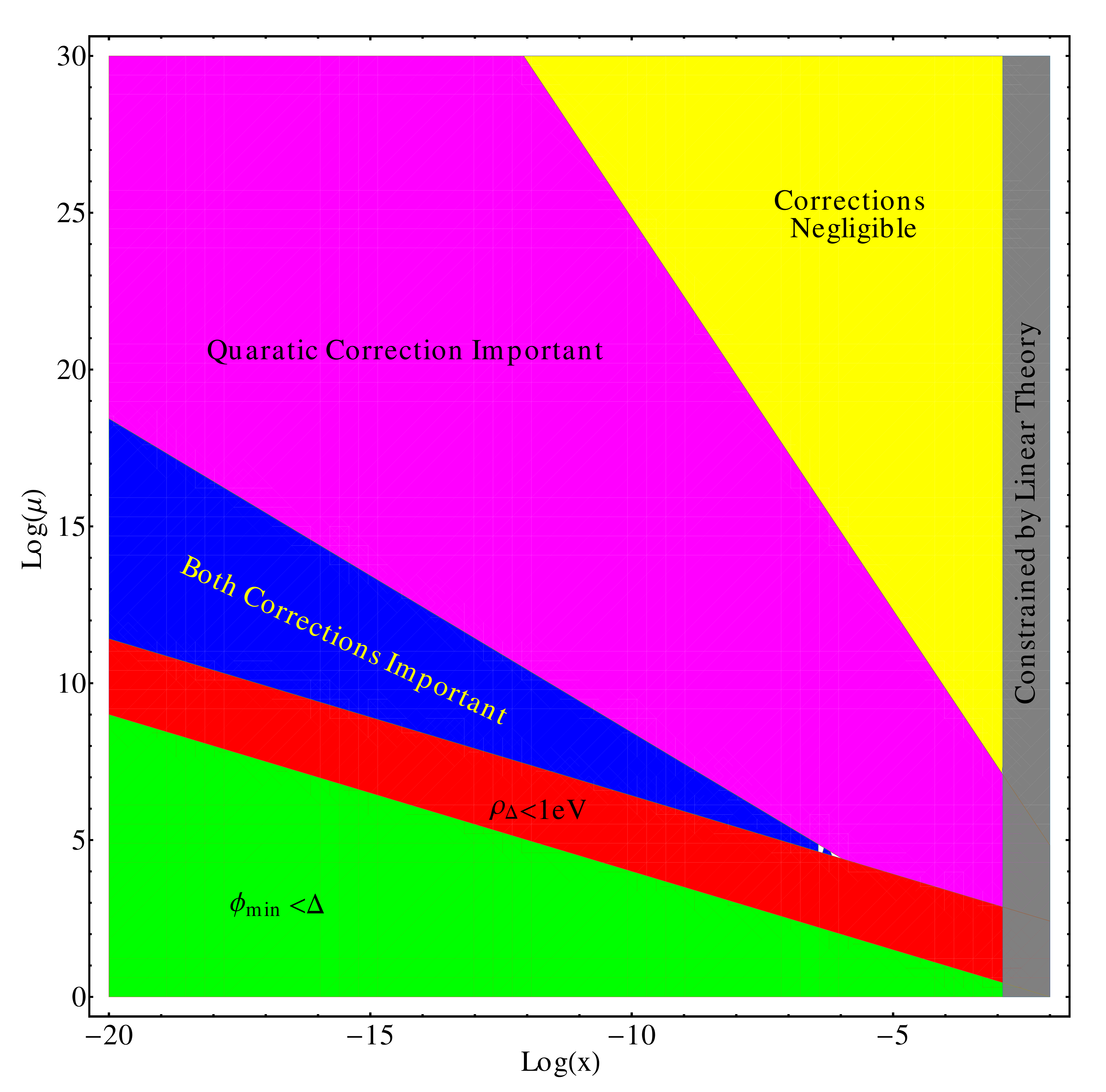}
\caption{The various regions in the $\log(\mu)$-$\log(x)$ plane with $n=\delta=2$, $z_\infty=5$ and $G=10^{-2}$. The yellow region shows the parameter range where the corrections are negligible. The magenta region shows the ranges where the quadratic correction is important and the dark blue region where both corrections are important. The red region corresponds to models where a cosmological constant is generated after last scattering and are therefore excluded and the green region corresponds to models where $\pmi<\Delta$ and a cosmological constant is only generated at some time in the future. The grey region corresponds to parameters where the model deviates from $\Lambda$CDM at the level of linear perturbations. }\label{fig:mx}
\end{figure}

\begin{figure}[ht]\centering
 \includegraphics[width=0.75\textwidth]{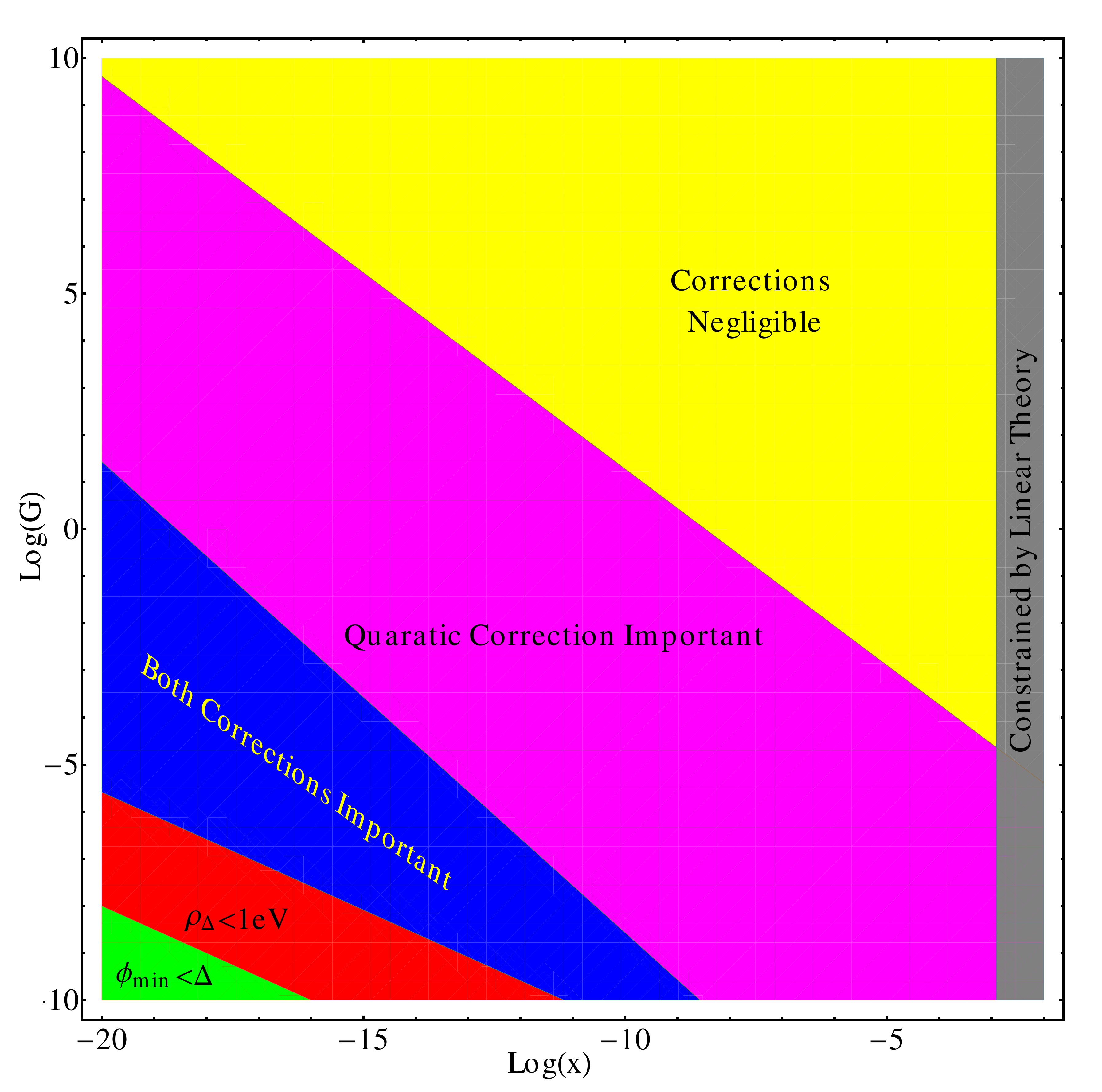}
\caption{The various regions in the $\log(G)$-$\log(x)$ plane with $n=\delta=2$, $z_\infty=5$ TeV and $\mu=10^{3}$ TeV. The yellow region shows the parameter range where the corrections are negligible. The magenta region shows the ranges where the quadratic correction is important and the dark blue region where both corrections are important. The red region corresponds to models where a cosmological constant is generated after last scattering and are therefore excluded and the green region corresponds to models where $\pmi<\Delta$ and a cosmological constant is only generated at some time in the future. The grey region corresponds to parameters where the model deviates from $\Lambda$CDM at the level of linear perturbations.}\label{fig:gx}
\end{figure}

Next, we plot the $\log(\mu)$-$\log(x)$ plane with $n=\delta=2$ and $G=10^{-2}$ in figure \ref{fig:mx} to investigate the effects of varying $\mu$ on the viable region. It is evident from the figure that large ($\gsim\mathcal{O}(\textrm{TeV})$) values of $\mu$ are required for there to be a large region with negligible corrections; in fact, if one steadily increases $\mu$ one finds that this region grows, replacing the regions where the corrections are important. This behaviour is traced back to equations (\ref{eq:rhoi}) and (\ref{eq:M4n}) in appendix C where it is shown that $M^{4+n}\propto \mu^n$ and therefore the density at which the corrections disappear increases slightly faster with $\mu$ than the densities at which the corrections become important.
Finally, now that we have some idea of the viable values of $z_\infty$ and $\mu$ we plot the $\log(G)$-$\log(x)$ plane with $n=\delta=2$ and $\mu=10^{3}$ TeV in figure \ref{fig:gx} in order to investigate the values of $G$ where the corrections are negligible. One can see that when $G\gsim\mathcal{O}(1)$ the corrections are generally negligible, which is a result of (\ref{eq:rhoi}) in appendix C, which show that the density at which the corrections disappear generally grows faster with $G$ than the density at which they become important. The density at which the corrections are important both include an explicit factor of $x^{-1}$ which is absent from the density at which the corrections disappear (there are other factors of $x$ coming from the scale $M$ though these vary with a far smaller power). This is the reason that the region where the corrections are negligible is larger when $x$ is closer to $1$. The plot clearly shows that there is a large region around $G\approx 10^{-2}$, which we have argued above is a sensible range where there is no excessive fine-tuning or strong $\mathrm{U}(1)$ coupling. With the parameters we have chosen this only exists when $x\gsim10^{-10}$, however this does not really constrain $x$. If one were to increase either $n$, $\delta$, $z_\infty$ or $\mu$ this region would extend further in the direction of decreasing $x$.
Thus, we have found that when the model parameters assume sensible values there is a large region of parameter space where the corrections to the effective potential are negligible before they vanish and an FI cosmological constant appears at late times. When the FI cosmological constant is generated, the background cosmology is indistinguishable from $\Lambda$CDM and so one must look at the growth of structure on linear scales in order to probe this theory. This will be examined in detail and here we show some of the results. The grey region in figures \ref{fig:zx}-\ref{fig:gx} correspond to parameters where the linear CDM power spectrum deviates from the GR prediction by up to $10$\% and can therefore be probed with upcoming experiments such as {\scriptsize EUCLID} \cite{Amendola:2012ys} whilst the other regions will show a negligible deviation from GR. Parameters that give deviations already ruled out by WiggleZ \cite{Contreras:2013bol} are not shown in the plots.

\section{Linear Perturbations}\label{sec:perts}

These models are always efficiently screened on small scales, however this screening mechanism is inherently non-linear and so one must still worry about fifth-force effects on the growth of structure on linear scales. It is entirely possible that the linear power spectrum is modified greatly in these models and it is important to impose constraints on the parameters such that the results of linear perturbation theory do not differ too much from GR. The growth of linear perturbations in screened modified gravity has been well studied \cite{Brax:2004qh,Brax:2005ew,Brax:2012gr} and the linear density contrast $\delta_{\rm c}=\delta\rc/\rc$ in the conformal Newtonian Gauge evolves on sub-horizon scales according to
\begin{equation}\label{eq:pertfull}
 \ddot{\dc}+2H\dot{\dc}-\frac{3}{2}\Omega_{\rm c}(a)H^2\left(1+\frac{2\beta^2(\varphi)}{1+\frac{m_{\rm eff}^2a^2}{k^2}}\right)\dc=0.
\end{equation}
The large mass of the field ensures that the perturbation $\delta\varphi(t,\vec{x})$ for $\varphi$ is given by \cite{Brax:2012gr}
\begin{equation}
\delta\varphi\approx-\frac{\beta(\varphi)}{k^2+a^2m_{\rm eff}^2}\frac{\rc}{\mpl}a^2\dc,
\end{equation}
which has been used to obtain the above equation. Physically, the last term in (\ref{eq:pertfull}) corresponds to a scale dependent enhancement of Newton's constant:
\begin{equation}\label{eq:effG}
 \frac{G_{\rm eff}(k)}{G}=1+\frac{2\beta(\varphi)^2}{1+\frac{a^2m_{\rm eff}^2}{k^2}}.
\end{equation}
On large scales the screening is effective and $G_{\rm eff}\approx G$ whilst on smaller scales the full enhancement, $G_{\rm eff}=G(1+2\beta(\varphi)^2)$ is felt. One would therefore expect that there is some wavenumber $\tilde{k}$, which we will calculate presently, below which the modes feel no significant fifth-forces and the GR power spectrum is recovered. Given (\ref{eq:minfbound}), we have $am_{3/2}\gg2.5\times10^{28}a\; \textrm{Mpc}^{-1}$ and so on the scales of interest we are always in the limit $k\ll m_\varphi a$. In this limit we can linearise (\ref{eq:pertfull}) to find
\begin{equation}\label{eq:pertlin}
 \ddot{\dc}+2H\dot{\dc}-\frac{3}{2}\Omega_{\rm c}(a)H^2\left(1+2k^2\frac{\beta^2(\varphi)}{m_{\rm eff}^2a^2}\right)\dc\approx0.
\end{equation}
Equation (\ref{eq:pertlin}) will be our starting point in what follows. It has solutions that can be written in terms of \textit{modified Bessel functions} (whose degree depends on the model parameters) and so we have provided a short introduction to these functions, including how equations of the form (\ref{eq:pertlin}) can be transformed into the modified Bessel equation, in appendix A. The quantity $\dc$ is gauge dependent and so is not a physical observable. Previous works have put constraints on their models by looking for the parameters where the final term in (\ref{eq:pertlin}) is small and the GR result is recovered. In fact, it is the linear power spectrum that is the physical observable and so here we shall study the deviations from the $\Lambda$CDM predictions as a potentially observable probe of supersymmetric chameleons.

In the radiation era, we have
\begin{equation}\label{eq:radpert}
 \ddot{\dc}+2H\dot{\dc}-\frac{3}{2}H^2 \left(\frac{G_{\rm eff}(k)}{G}\right)\frac{\bar{\rc}}{\bar{\rho_c}+\bar{\rho_{\rm r}}}\dc=0,
\end{equation}
where $G_{\rm eff}(k)$ is defined in \ref{eq:effG} above and $\bar{\rc}/(\bar{\rho_c}+\bar{\rho_{\rm r}})\approx(1+z_{\rm eq})/(1+z)$. In GR, $G_{\rm eff}=G$ and this final term is negligible compared to the time derivatives, which scale as $H^2$, and can be neglected to give a logarithmic growth of the density contrast, $\dc\propto\ln(t)$. With the inclusion of modified gravity, one has
\begin{equation}\label{eq:radGeff}
\frac{G_{\rm eff}-G}{G}\left(\frac{1+z_{\rm eq}}{1+z}\right)\approx \mathcal{O}(1)\,x\left(\frac{k}{10^{-5}\textrm{ Mpc}^{-1}}\right)^2\frac{1}{(1+z)^{\frac{2n+5\delta}{n+\delta}}},
\end{equation}
where $H_0^2\sim 10^{-5}h^2$ Mpc$^{-1}$. Now modes deep inside the horizon can have arbitrarily large values of $k$, however, we know that, in GR at least, modes with $k>0.1h$ Mpc$^{-1}$ will be non-linear today even if they were not in the past so that we do not need to keep track of their evolution. In modified gravity, we expect this number to be smaller, however here we will use the GR value in order to be conservative. In this case, the largest mode which is linear today satisfies $k/10^{-5}h\textrm{ Mpc}^{-1}\sim10^4$.
Now $z\gg z_{\rm eq}\sim 10^3$ and the minimum value of $(2n+5\delta)/(n+\delta)$ is 2 so that the maximum deviation from GR satisfies
\begin{equation}
\left.\frac{G_{\rm eff}-G}{G}\left(\frac{1+z_{\rm eq}}{1+z}\right)\right\vert_{\textrm{max}}\ll 10^2x.
\end{equation}
Unless $x$ assumes unrealistically large values $x\sim\mathcal{O}(1)$, which we shall see below gives large deviations from the CDM power spectrum today in tension with current observations, the final term in \ref{eq:radpert} is negligible and the modes evolve in an identical manner to GR.

We can use the standard GR result and, for simplicity, will not use the full logarithmic form but will treat the modes as constant inside the radiation era\footnote{This approximation may be relaxed with little effort and indeed should be if one wished to compare to data, however, given that we shall not do so here there is little to be gained by including the logarithmic term.}. Outside the horizon, both during the matter and radiation eras, we have $G_{\rm eff}\approx G$ since the modifications of GR are suppressed and so the perturbations do not evolve. We hence treat the modes as constant until the time of horizon re-entry. Finally, we shall treat the change from radiation to matter domination as a sharp transition. Whilst not strictly necessary, this allows us to compute the power spectrum in closed form and there are no subtleties associated with modified gravity in treating the full transition period. We make this approximation in the interest of discerning the new features introduced by supersymmetric chameleons as simply as possible. 
 
On the scalar field side, we assume that the field settles into its supersymmetric minimum instantaneously at $\rc=\rho_\infty$, thereby ignoring the short-lived transition period when equation (\ref{eq:phimineq}) has no closed-form solution and any oscillations around the minimum. This was studied in a specific case in \cite{Brax:2004qh}, where it was found that the sharp settling is a very good approximation and so we do not expect the short-lived transition period to impact upon the power spectrum. Our power spectrum therefore exhibits unphysical sharp discontinuities at the scales which enter the horizon at matter-radiation equality and (as we shall see momentarily) at $z_\infty$, which would be found to be smooth curves had we solved the full equations numerically. We will primarily be concerned with the power spectrum on large scales, since this is where the power spectrum is most unconstrained and so these unphysical features will play no part in our conclusions.

In what follows we shall consider two distinct cases. In the previous subsection we found that we can account for dark energy without a cosmological constant by imposing $z_\infty<0$, $\rc\gg\rho_\infty$ and $\delta\gg n $ and so we shall first investigate this case. We then go on to investigate the general case $z_\infty\gsim0$ where it is assumed that the field reaches its supersymmetric minimum sometime around the current epoch i.e. $z_\infty\lsim10$ although our treatment will be valid for $z_\infty\le z_{\rm eq}$.

In the following we focus on the power spectrum in the matter era. After the end of the matter era around a redshift $z\sim 1$, the growth of structure is slowed down by the presence of dark energy. Hence deviations of the power spectrum from $\Lambda$-CDM are maximally dependent on the features of modified gravity when calculated at the end of the matter era. This can be extended to later times by numerical calculations which are left to future work. Here we shall only make sure that the deviations from GR are no larger than around 10\% to comply with recent observations \cite{Contreras:2013bol}. In the future, large scale surveys like {\scriptsize EUCLID} will test the linear at the percent level \cite{Amendola:2012ys} putting more constraints on super-chameleon models. The results of these calculations have been used in figures \ref{fig:zx}-\ref{fig:gx} to indicate where exactly in the parameter space these deviations occur.

\subsection{$z_\infty<0$}

We have seen in section \ref{sec:backcos} that when $\delta\gg n$ and $x\delta(1+z_\infty)^3\approx3n\Omega_\Lambda^0$ we can have $w_\varphi\approx-1$ and $V_{\rm eff}\sim \mpl^2H_0^2$, consistent with the current dark energy observations. Since this case is of particular interest to us we will enforce these conditions below and refer to them as the \textit{dark energy conditions}. Assuming a matter dominated era we can use equations (\ref{eq:betaphi}) and (\ref{eq:mphi}) in (\ref{eq:pertlin}) to find
\begin{equation}\label{eq:pertvia}
 t^2\ddot{\dc}+\frac{4}{3}t\dot{\dc}-\left[\frac{2}{3}\Omega_{\rm c}^0+9x(1+z_\infty)^3(kt_0)^2\left(\frac{t}{t_0}\right)^{\frac{8}{3}}\right]\dc.
\end{equation}
Following appendix A, the growing mode solution is
\begin{equation}\label{eq:d>npert}
 \dc(t)=C_{\rm MG}(k)t^{-\frac{1}{6}}I_\nu\left[\sigma kt_0\left(\frac{t}{t_0}\right)^{\frac{4}{3}}\right],
\end{equation}
where
\begin{equation}
 \nu^2=\frac{1}{8}\left(\frac{1}{8}+3\Omega_{\rm c}\right)\quad\textrm{and}\quad \sigma^2\equiv \frac{81x(1+z_\infty)^3}{16}\approx \frac{243}{16}\frac{n\Omega_\Lambda^0}{\delta},
\end{equation}
where the last equality for $\sigma$ holds when we impose the conditions that the field to account for dark energy. This should be compared with the GR prediction
\begin{equation}
 \dc(t)=C_{\rm GR}(K)t^{n};\quad\textrm{where}\quad n=-\frac{1}{6}+\frac{1}{2}\sqrt{\frac{1}{9}+\frac{8}{3}\Omega_{\rm c}^0},
\end{equation}
where $\Omega_{\rm c}^0\sim 1$ at the end of the matter era.
For small $x$, we have $I_\nu(x)\sim x^\nu\left[1+\mathcal{O}(x^2)\right]$ to leading order (see appendix A) and noting that $8\nu=6n+1$ we can see that these expressions agree for small $k$. Given the solution (\ref{eq:d>npert}) we are now in a position to calculate the power spectrum. We start by noting that the time at which a given mode crosses the horizon ($k=2\pi a H$) is
\begin{equation}
 t_{\rm H}=t_0\left(\frac{4\pi}{3t_0k}\right)^3.
\end{equation}
and assume that the modes are constant during the radiation era and outside the horizon in the matter era as discussed above. In this case, the contrast during the radiation era (and outside the horizon in the matter era) is given by the primordial fluctuations from inflation, $\dc^{\rm I}$. Modes that enter the horizon during the matter era, that is modes with $k<k_{\rm eq}=0.01 h$ Mpc$^{-1}$, will begin evolving according to (\ref{eq:d>npert}) and so we have the boundary condition $\dc(\h)=C_{\rm MG}(k)\h^{-1/6}I_\nu[\sigma kt_0(\h/t_0)^{4/3}]$, which allows us to find $C_{\rm MG}(k)$ and hence the power spectrum
\begin{equation}\label{eq:ps_dn}
 P(k)=\langle\left\vert\dc(t_0)\right\vert^2\rangle=\langle\left\vert\dc^{\rm I}\right\vert^2\rangle\left\{\begin{array}{l l}
\left(\frac{\h}{t_0}\right)^{\frac{1}{3}}\frac{I^2_\nu\left(\sigma kt_0\right)}{I^2_\nu\left[\sigma kt_0\left(\frac{\h}{t_0}\right)^{\frac{4}{3}}\right]}& k<k_{\rm eq}\\
\left(\frac{t_{\rm eq}}{t_0}\right)^{\frac{1}{3}}\frac{I^2_\nu\left(\sigma kt_0\right)}{I^2_\nu\left[\sigma kt_0\left(\frac{t_{\rm eq}}{t_0}\right)^{\frac{4}{3}}\right]}& k>k_{\rm eq}
\end{array}\right. .
\end{equation}
Modified Bessel functions diverge from their leading order GR behaviour very rapidly and so if the modified power spectrum is not to deviate from the GR prediction too greatly the arguments of both functions in (\ref{eq:ps_dn}) must be small enough such that the leading order behaviour is a good approximation, at least over the entire range of $k$ where linear theory is valid. Since $t_0>\h$ this is equivalent to demanding $\sigma k t_0\ll1$. When this is satisfied the power spectrum will show no deviations from GR and when this begins to break down we expect to see small deviations. One can verify using equation (\ref{eq:seriesI}) in appendix A that the leading order dependence is $k^4$, the same as predicted by GR. Alternatively, according to (\ref{eq:effG}) we should recover the GR prediction whenever $k\ll m_{\rm eff}/\beta(\varphi)$. Using equations (\ref{eq:betaphi}) and (\ref{eq:mphi}) one finds an equivalent condition up to an order unity coefficient. 

One can then define the above-mentioned scale, $\tilde{k}$, below which no gravitational enhancement is felt:
\begin{equation}
 \tilde{k}^2\simeq \frac{H_0^2}{x(1+z_\infty)^3}=\frac{\delta H_0^2}{3n\Omega_\Lambda^0},
\end{equation}
where the last equality holds when we impose the dark energy conditions. Since $k\sim H_0$ corresponds to modes which enter the horizon today we expect deviations from GR on smaller scales. Unless $\delta\gg10^6n$, the linear CDM power spectrum differs from the GR one by several orders of magnitude. Since $\delta$ appears as an index in the superpotential such a large value seems highly unnatural. With this in mind, we will abandon this limit and proceed to study the general case where we allow $n$ and $\delta$ to vary independently and $\varphi$ to converge to its supersymmetric minimum at some point in the recent past.

\subsection{$z_\infty>0$}\label{sec:pertg}

We start by noting that when $z_\infty>0$ the final term in (\ref{eq:pertlin}) will exhibit a different time dependence after the field has converged to its supersymmetric minimum, so we must keep track of modes that enter the horizon before and after this and match the time evolution appropriately. We therefore begin by solving (\ref{eq:pertlin}) for the case where $z<z_\infty$ so that $\varphi\approx\varphi_{\rm min}$ and $m_\varphi\approx m_\infty$. Assuming a matter dominated epoch and defining $k_\infty=2\pi a(t_\infty)H(t_\infty)$ to be the mode which enters the horizon when $z=z_\infty$ ($t_\infty=t_0(1+z_\infty)^{-3/2}$) we have
\begin{equation}\label{eq:pertzlzi}
 t^2\ddot{\dc}+\frac{4}{3}t\dot{\dc}-\left[\frac{2}{3}\Omega_{\rm c}^0+\frac{9x\delta}{n(1+z_\infty)^{3}}(kt_0)^2\left(\frac{t_0}{t}\right)^{\frac{4}{3}}\right]\dc\quad k<k_\infty.
\end{equation}
Following appendix A we can again write down the solution in terms of modified Bessel functions. Since the final term decreases with increasing $t$ the growing mode is the modified Bessel function of the second kind:
\begin{equation}\label{eq:infty}
 \dc(t)=C_{\rm MG}^{k<k_\infty}(k)t^{-\frac{1}{6}}K_\omega\left[\zeta kt_0\left(\frac{t_0}{t}\right)^{\frac{2}{3}}\right],
\end{equation}
where
\begin{equation}\label{eq:zeta}
 \zeta^2\equiv \frac{27x\delta}{2n(1+z_\infty)^3};\quad\textrm{and}\quad \omega^2=\frac{9}{4}(\frac{1}{36}+\frac{2}{3}\Omega_{\rm c}^0).
\end{equation}
Next, we must find the solution when $k>k_\infty$. In this case we have (using equations (\ref{eq:betaphi}), (\ref{eq:vphimin}) and (\ref{eq:mphimin}))
\begin{equation}\label{eq:pertgenzg0}
 t^2\ddot{\dc}+\frac{4}{3}t\dot{\dc}-\left[\frac{2}{3}\Omega_{\rm c}^0+9\frac{x\delta}{n+\delta}(1+z_\infty)^{\frac{3\delta}{n+\delta}}k^2t_0^2\left(\frac{t}{t_0}\right)^{\frac{8\delta+2n}{3(\delta+n)}}\right]\dc,
\end{equation}
the solution of which is
\begin{equation}\label{eq:generalpert}
 \dc(t)=C_{\rm MG}^{k>k_\infty}(k)I_{\nu}\left[\sigma kt_0\left(\frac{t}{t_0}\right)^r\,\right];\quad \nu^2=\left(\frac{\delta+n}{4\delta+n}\right)^{2}\left[\frac{1}{4}+6\Omega_{\rm c}^0\right],
\end{equation}
with
\begin{equation}
 \sigma^2\equiv\frac{81x\delta(\delta+n)(1+z_\infty)^{\frac{3\delta}{\delta+n}}}{(4\delta+n)^2}\quad\textrm{and}\quad r=\frac{4\delta+n}{3(\delta+n)}.
\end{equation}
We can use this to calculate the power spectrum in the general case. Modes that enter the horizon during the radiation dominated era (i.e. $k>k_{\rm eq}$) are constant until matter radiation equality when they start growing according to equation (\ref{eq:generalpert}). In this case we have $C_{\rm MG}^{k>k_\infty}(k>k_{\rm eq})I_{\nu}[\sigma(t_{\rm eq}/t_k)^r]=\dc(t_{\rm eq})$. On the other hand, modes which enter during matter domination are subject to the condition $C_{\rm MG}^{k>k_\infty}(k<k_{\rm eq})I_{\nu}\left[\sigma(\h/t_k)^r\right]=\dc(\h)$. Thus, modes that enter the horizon before $z_\infty$ evolve according to
\begin{equation}
 \dc(t)= \dc^{\rm I}\left\{
 \begin{array}{l l}
  \left(\frac{t_{\rm eq}}{t}\right)^{\frac{1}{6}}\frac{I_{\nu}\left[\sigma kt_0\left(\frac{t}{t_0}\right)^r\right]}{I_{\nu}\left[\sigma kt_0\left(\frac{t_{\rm eq}}{t_0}\right)^r\right]}& k>k_{\rm eq} \\
  \left(\frac{\h}{t}\right)^{\frac{1}{6}}\frac{I_{\nu}\left[\sigma kt\left(\frac{t}{t_0}\right)^r\right]}{I_{\nu}\left[\sigma kt_0\left(\frac{\h}{t_0}\right)^r\right]}&k<k_{\rm eq}\\
 \end{array}\right. .
\end{equation}
Near $z_\infty$, modes inside the horizon (whether they entered during the radiation or matter era) thus evolve according to the general form $\dc(t)=D(k)t^{-1/6}I_\nu[\sigma kt_0 (t/t_0)^{r}]$ where the form of $D(k)$ varies depending on when the mode entered the horizon as detailed above. When $z=z_\infty$ the field converges to its supersymmetric minimum and the evolution now proceeds according to equation (\ref{eq:infty}) and we must again match the two solutions at $z=z_\infty$ so that $\dc(t_\infty)=D(k)t_\infty^{-1/6}I_\nu[\sigma kt_0 (t_\infty/t_0)^{r}].$ Modes that enter the horizon later than this simply evolve according to (\ref{eq:infty}), matching at the time when they enter the horizon, in which case we have $\dc(\h)=C_{\rm MG}^{k<k_\infty}(k)\h^{-1/6}K_\omega[\zeta kt_0(t_0/\h)^{2/3}]$. This leaves the power spectrum taking on different functional forms in three different ranges of $k$ (this is to be contrasted with the two predicted in GR):
\begin{equation}\label{eq:genps}
 P(k)=\left\langle\left\vert\dc^{\rm I}\right\vert^2\right\rangle\left\{
 \begin{array}{l l}
  \left(\frac{t_{\rm eq}}{t_0}\right)^{\frac{1}{3}}\frac{I^2_{\nu}\left[\sigma kt_0\left(\frac{t_\infty}{t_0}\right)^r\right]}{I^2_{\nu}\left[\sigma kt_0\left(\frac{t_{\rm eq}}{t_0}\right)^r\right]}\frac{K_\omega^2\left[\zeta kt_0\right]}{K^2_\omega\left[{\zeta kt_0}{(1+z_\infty)}\right]}& k>k_{\rm eq}>k_\infty \\
  \left(\frac{\h}{t_0}\right)^{\frac{1}{3}}\frac{I^2_{\nu}\left[\sigma kt_0\left(\frac{t_\infty}{t_0}\right)^r\right]}{I^2_{\nu}\left[\sigma kt_0\left(\frac{\h}{t_0}\right)^r\right]}\frac{K_\omega^2\left[\zeta kt_0\right]}{K^2_\omega\left[{\zeta kt_0}{(1+z_\infty)}\right]}&k_\infty<k<k_{\rm eq}\\ \left(\frac{\h}{t_0}\right)^{
\frac{1}{3}}\frac{K^2_\omega\left[\zeta kt_0\right]}{K_\omega^2\left[\zeta kt_0\left(\frac{t_0}{\h}\right)^{\frac{2}{3}}\right]}
&k<k_\infty
 \end{array}\right. .
\end{equation}

We are now in a position to explore the deviations from the GR prediction, but we must first check that GR is indeed recovered on large scales. We know from our analysis in the previous subsection that this requires taking the argument of all modified Bessel functions of the first kind to be small, however the second kind functions require more thought. As detailed in appendix A, these grow with decreasing argument and diverge as it approaches zero and so one may be concerned that taking the argument to be small is not the correct limit. In fact, $K_\omega[y]\sim y^{-\omega}+\mathcal{O}(y^{2-\omega})$ (see appendix A) and so in this limit one may neglect the higher order terms. One can indeed check by expanding the functions according to (\ref{eq:seriesI}) that this leading order behaviour coincides with the GR prediction.

\begin{figure}[ht]\centering
 \includegraphics[width=0.75\textwidth]{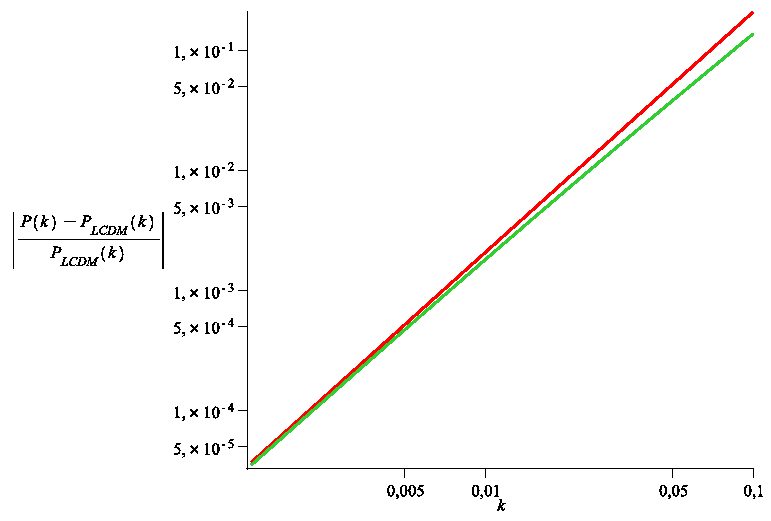}
\caption{The deviation from the $\Lambda$CDM power spectrum when a cosmological constant is present, with $\delta=n=2$, $x=5\times 10^{-7}$ and $z_\infty=5$. In red we have represented the exact power spectrum deviation and in green its linearisation. Lower values of $x$ give even less power spectrum deviation.}
\end{figure}

One can then go beyond leading order to find the predicted deviations from $\Lambda$CDM. Modified Bessel functions deviate from their leading order terms very rapidly and so one is interested in the case where the deviations are given by the next-to-leading order expansion. In this regime, the power spectrum \ref{eq:genps} can be expanded (using the power series given
in appendix A) and deviates from the $\Lambda$CDM case by $\Delta P(k)$, which
is scale dependent:
\begin{equation}\label{eq:genps2}
 \frac{\Delta P(k)}{P_{\rm \Lambda CDM}(k)}=\left\{  \begin{array}{l l}
   \frac{2}{\nu +1} \left(\frac{\sigma k t_0}{2}\right)^{2r} \left(\frac{t_\infty}{t_0}\right)^{2r}\left[1-\left(\frac{t_{eq}}{t_\infty}\right)^{2r}\right] +2F(\omega)\left(\frac{\zeta k t_0}{2}\right)^2\left[(1+z_\infty)^2-1\right] & k>k_{\rm eq}>k_\infty \\      \frac{2}{\nu +1} \left(\frac{\sigma k t_0}{2}\right)^{2r} \left(\frac{t_\infty}{t_0}\right)^{2r}\left[1-\left(\frac{\h}{t_\infty}\right)^ {2r}\right] +2F(\omega)\left(\frac{\zeta k t_0}{2}\right)^2\left[(1+z_\infty)^2 -1\right]&k_\infty<k<k_{\rm eq}\\ 2F(\omega)\left(\frac{\zeta k t_0}{2}\right)^2 \left[\left(\frac{t_0}{\h}\right)^{4/3}-1\right]
&k<k_\infty
  \end{array}\right. .
\end{equation}
The factor $F(\omega)$ is given by
\begin{equation}\label{eq:F(w)}
 F(\omega)= \left\{
  \begin{array}{l l}
   \frac{1}{\omega-1}& \omega>1 \\
    \frac{\pi\omega}{\Gamma^2(1+\omega)\sin(\pi\omega)}&\omega<1\\
  \end{array}\right. .
\end{equation}
The case $\omega=1$ corresponds to $\Omega_{\rm c}^0=0.625$, which is not
physically relevant and so is not considered here. This factor arises as a
result of different terms in the expansion of $K_\omega[y]$ becoming sub-leading
when $\omega$ assumes different values. The technical details are given in
appendix A.

The deviation monotonically increases with $k$ in the linear regime. On large
scales entering the horizon after $t_\infty$, the growth of the deviation
$\Delta P/P_{\rm \Lambda CDM}$ is in $k^6$ whereas for smaller scales entering the
horizon before $t_\infty$, the discrepancy grows as $k^2$. Of course, this
result is only valid in the linear regime of perturbation theory. On smaller scales,
typically $k\gsim 0.1 {\rm hMpc}^{-1}$, non-linear effects become important and the screening mechanism is active. We therefore expect that the deviations
from $\Lambda$CDM will be rapidly suppressed. As a result, we expect the full power spectrum to show
deviations from $\Lambda$CDM in the linear regime only and none at all in
the non-linear one. Of course the complete analysis of this phenomenon may
require N-body simulations which are beyond the scope of the present paper.

The CDM power spectrum predicted by the super-chameleon is very different from its
chameleonic counterpart as exemplified by $f(R)$ models in the large curvature
regime. Indeed, in this case linear scales are slightly outside
the Compton wavelength of the scalar field whose range is around 1 Mpc.
Deviations from $\Lambda$CDM are small until the quasi-linear regime for scales
around 1 Mpc where the effects of the scalar peaks before being damped in the
non-linear regime by the screening mechanism. Here, scales are always outside
the Compton wavelength of the scalar field and a deviation from $\Lambda$CDM
is only possible due to the large coupling to matter. This can be
effective enough to lead to $k$ dependent effects in the linear regime before
being heavily Yukawa suppressed in the non-linear one. Hence we have found that
super-chameleons leave a drastically different signature on large scale
structure formation than the majority of previously studied models.

\section{Conclusions}\label{sec:concs}

We have presented a class of globally supersymmetric chameleons coupled to dark matter fermions and have investigated their cosmological dynamics. Like all supersymmetric chameleons, except those arising from no-scale K\"{a}hler potentials, these models exhibit such efficient screening that there are no astrophysical signatures\footnote{We have not investigated a coupling to photons, which is not bound by these general results.}. At small field values the scalar potential is locally run-away, however, at larger values this behaviour terminates at a supersymmetric minimum. The coupling to fermions results in a density dependent effective potential and at finite CDM densities the field is displaced from this minimum to smaller values, breaking supersymmetry. Since the dark sector is secluded from the matter sector the scale of this breaking depends only on the model parameters and the ambient density so the scalar field's VEV does not receive TEV-scale corrections coming from the observable sector.
In order to address the issue of a cosmological constant, we have made use of a mechanism introduced in a previous letter \cite{Brax:2012mq}. A coupling of the super-chameleon to two scalars charged under a local $\mathrm{U}(1)$ symmetry results in a chameleon-dependent effective mass for one of the scalars. At early times this mass is negative, however as the cosmological super-chameleon evolves to larger values this steadily increases until it reaches zero, restoring the symmetry. When this happens, only a Fayet-Illiopoulos term remains in the scalar potential, which acts as a cosmological constant. If we assume that the old cosmological constant problem in the observable sector is solved and tune this FI term we can reproduce the small cosmological constant needed to match current observations. Our low-energy description does not include a natural value for this FI term, indeed it is arbitrary, however it is robust against quantum corrections and so if one could find a more UV complete model where there is a natural mechanism explaining this tuning then this value would survive to arbitrarily low energy scales.

The mass of the field is at least as large as the gravitino mass and so the density-dependent minimum of this effective potential is a stable attractor, which the cosmological field tracks from small values at early times to the supersymmetric minimum at late times. We have examined the growth of linear structures in these theories and, by solving the modified equation for the time-evolution of the density contrast in closed form, have derived the modified CDM power spectrum. Interestingly, this modified power spectrum exhibits three distinct scale variations rather than the two predicted by GR. Moreover, for a large choice of model parameter, the deviations of the power spectrum in the linear regime has a characteristic $k^2$ growth for scales entering the horizon before matter-radiation equality.

The efficient screening precludes all astrophysical signatures and the background cosmology is indistinguishable from $\Lambda$CDM but we have also found that small deviations of the power spectrum are present with characteristic features for a large part of the parameter space and therefore the models can be falsified with upcoming experiments. It would be useful to investigate this further in the quasi-linear to the non-linear regimes where the screening effects should play a fundamental role. This is left for future work and here we shall be content that we can find supersymmetric models of screened modified gravity with characteristic deviations from $\Lambda$CDM at the linear level.

\section*{Acknowledgements}

We are grateful to Neil Barnaby, Daniel Baumann, Eugene Lim and David Marsh for helpful discussions. JS and ACD (in part) is supported by the STFC.
\appendix
\section{Modified Bessel Equations}
\subsection{Generalised Modified Bessel Equations}
We have seen in section \ref{sec:perts} that the linearised perturbation equations are all of the form
\begin{equation}
 t^2\ddot{\dc}+at\dot{\dc}-\left(b^2+c^2t^{2r}\right)\dc=0,
\end{equation}
with $a=4/3$. The substitution $\dc=t^{n}\tdc(t)$ with $n=(1-a)/2$ may be used to find the following equation for $\tdc$:
\begin{equation}
 t^2\ddot{\tdc}+t\dot{\tdc}-\left(\frac{(a-1)^2}{4}+b^2+c^2t^{2r}\right)\tdc=0,
\end{equation}
which may further be transformed into the form
\begin{equation}\label{eq:modbes}
 u^2\tdc^{\prime\prime}+u\tdc^\prime-\left(\nu^2+u^2\right)\tdc;\quad \nu^2\equiv \frac{(a-1)^2}{4r^2}+\frac{b^2}{r^2}
\end{equation}
using the substitution $u=ct^r/r$ and the notation $\prime\equiv \dd/\dd u$. Equation (\ref{eq:modbes}) is a \textit{modified Bessel equation}, the solutions of which are \textit{modified Bessel functions} of the first and second kind, $I_\nu(u)$ and $K_\nu(u)$. Unlike regular Bessel, functions which are oscillatory in nature, these functions either grow ($I_\nu$) or decay ($K_\nu$) with increasing $u$. The general solution is then
\begin{equation}
 \dc(t)=t^{\frac{1-a}{2}}\left[C_1I_\nu\left(\frac{c}{r}t^r\right)+C_2K_\nu\left(\frac{c}{r}t^r\right)\right]
\end{equation}
although in section \ref{sec:perts} we shall only be interested in whichever function is the growing mode (this depends on the sign of $r$). The modified Bessel function of the first kind has the power series expansion
\begin{equation}\label{eq:seriesI}
 I_\nu(x)=\sum_{k=0}^\infty\frac{1}{k!\Gamma(\nu+k+1)}\left(\frac{x}{2}\right)^{\nu+2k}=\frac{1}{\Gamma(1+\nu)}\left(\frac{x}{2}\right)^\nu\left(1+\mathcal{O}(x^2)+\ldots\right),
\end{equation}
where $\Gamma(m)$ is the gamma function. The modified Bessel function of the second kind is defined via
\begin{equation}
 K_\nu(x)=\lim_{n\to \nu}\frac{\pi\left[I_{-n}(x)-I_n(x)\right]}{2\sin(n\pi)}.
\end{equation}
Its power series expansion for $\nu\notin\mathbb{Z}$ is as follows:
\begin{equation}\label{eq:seriesK}
 K_{\nu}(x)=\frac{\pi\textrm{csc}(\pi\nu)}{2}\left[\sum_{k=0}^\infty\frac{1}{
\Gamma(k-\nu+1)k!}\left(\frac{x}{2}\right)^{2k-\nu}-\sum_{k=0}^\infty\frac{1}{
\Gamma(k+\nu+1)k!}\left(\frac{x}{2}\right)^{2k+\nu}\right] .
\end{equation}
\subsection{Second Kind Power Series Expansions for the Power Spectra}
In this section we briefly outline the steps that lead to the factor
$F(\omega)$ given in \ref{eq:F(w)} and appearing in the deviation from the
$\Lambda$CDM spectra \ref{eq:genps2}. The expansion of the first kind modified
Bessel functions \ref{eq:seriesI} is a trivial exercise in algebra, however
the second kind expansion \ref{eq:seriesK} requires more thought. The three
leading terms are:
\begin{equation}
 K_{\nu}[x]=\frac{\pi\textrm{csc}(\pi\nu)}{2\Gamma(1-\nu)}\left(\frac{2}{x}
\right)^\nu\left[1+\frac{\Gamma(1-\nu)}{\Gamma(2-\nu)}\left(\frac{x}{
2}\right)^2-\frac{\Gamma(1-\nu)}{\Gamma(1+\nu)}\left(\frac{x}{2}\right)^{2\nu}
\right],
\end{equation}
which correspond to the first two terms in the first sum in \ref{eq:seriesK}
and the first term in the second sum. The next-to-leading order correction to
the leading order term ($\propto x^{-\nu}$) depends on whether $\nu>1$ or
the converse, which gives rise to the factor $F(\omega)$ in \ref{eq:genps2}.
When $\nu=1$ a different power series is needed, which we do not give here
since it corresponds to an uninteresting scenario. When $\nu<1$, the
final factor can be evaluated using the relation
\begin{equation}
 \Gamma(1-m)\Gamma(m)=\frac{\pi}{\sin(\pi m)}.
\end{equation}

\section{Minimisation of the Global Potential}
In this appendix we show how the global F-term scalar potential can be minimised to eliminate both the angular fields and $\pi_-$ to recover the simple form given in equation (\ref{eq:Fpot}). Ignoring the contribution from the K\"{a}hler potential for now (it depends on $|\phi|$ only) and setting $\langle\phi_\pm\rangle=0$, which is always a minimum, we have:
\begin{align}\label{eq:dW}
 \left\vert\frac{\dd
W}{\dd\phi}\right\vert^2=&{g^\prime}^2|\pi_+|^2|\pi_-|^2+\frac{\gamma^2}{2}
\left(\frac{|\phi|^{2\alpha-2}}{\Lambda_0^{2\alpha-6}}+\frac{|\phi|^{2\gamma-2}}
{ \Lambda_2^{2\gamma-6}}\right)
+\gamma\Re\left(\frac{\phi^{\alpha-1}{\phi^*}^{\gamma-1}}{\Lambda_0^{\alpha-3}
\Lambda_2^{\gamma-3}}\right)\nonumber\\&+\sqrt{2}{g^\prime}\gamma\Re\left[
\pi_+\pi_-\left(\frac{{\phi^*}^{\alpha-1}}{\Lambda_0^{\alpha-3}}+\frac{{\phi^*}^
{\gamma-1}}{\Lambda_2^{\gamma-3}}\right)\right],
\end{align}
which, as can be seen, simplifies greatly when the negatively charged field has zero VEV. We shall see now that this VEV does indeed minimise the potential. We begin by writing the charged fields in polar form $\pi_\pm\equiv \pi_\pm e^{i\theta_\pm}$. Ideally, one would hope to set the three angular fields $\{\theta,\theta_\pm\}$ to constant values in order to give negative signs in from of the final three terms in (\ref{eq:dW})\footnote{This is the approach taken when the charged fields are absent \cite{Brax:2011qs}.}, however this is not possible and instead one must eliminate them in terms of the other fields. In order to do this, we exploit the local $\mathrm{U}(1)$ symmetry, which acts as $\theta_\pm(x)\rightarrow\pm q\alpha(x)$, to set $\theta_+(x)=\theta_-(x)=\chi(x)/2$, which reduces the angular fields to the set $\{\theta,\chi\}$. With this in mind, the scalar potential, including the contribution from the K\"{a}hler potential is
\begin{align}\label{eq:global_V}
 V(\phi,&\theta,\pi_+,\pi_-,\chi)=\frac{2{g^\prime}^2\pi_+^2\pi_-^2}{\gamma^2}
\left(\frac{\Lambda_1}{\phi}\right)^{2\gamma-2}+\left(\frac{\Lambda_1}{\phi}
\right)^{2\gamma-2}\left(\frac{\phi^{2\alpha-2}}{\Lambda_0^{2\alpha-6}}+\frac{
\phi^{2\gamma-2}}{\Lambda_2^{2\gamma-6}}\right)\nonumber\\&
+\frac{2}{\gamma}\left(\frac{\Lambda_1}{\phi}\right)^{2\gamma-2}\frac{\phi^{
\alpha+\gamma-2}}{\Lambda_0^{\alpha-3}\Lambda_2^{\gamma-3}}\cos[
(\alpha-\gamma)\theta]\nonumber\\&+\frac{g^\prime}{\sqrt{2}\gamma}\left(\frac{
\Lambda_1}{\phi}\right)^{2\gamma-2}\pi_+\pi_-\left[\frac{\phi^{\alpha-1}}{
\Lambda_0^{\alpha-3}}\cos[\chi-(\alpha-1)\theta]+\frac{\phi^{\gamma-1}}{
\Lambda_2^{\gamma-3}}\cos[\chi-(\gamma-1)\theta]\right].
\end{align}
Minimising this with respect to $\chi$ one finds
\begin{equation}
 \frac{\phi^{\alpha-\gamma}}{\Lambda_0^{\alpha-3}}\sin[\chi-(\alpha-1)\theta]
+\frac{1}{\Lambda_2^{\gamma-3}}\sin[\chi-(\gamma-1)\theta]=0,
\end{equation}
which may be used in the equation found by minimising equation
(\ref{eq:global_V}) with respect to $\theta$ to find a relation between
$\sin[(\alpha-\gamma)\theta]$ and $\sin[\chi-(\gamma-1)\theta)]$ (or
equivalently $\sin[\chi-(\alpha-1)\theta]$):
\begin{equation}\label{eq:sinsin}
 \frac{2\phi^{\alpha-1}}{\Lambda_0^{\alpha-3}}\sin[(\alpha-\gamma)\theta]=\frac{
g^\prime}{2}\pi_+\pi_-\sin[\chi-(\gamma-1)\theta].
\end{equation}
This may be used to eliminate $\chi$ from the potential to find (with $\psi$ fixed at its minimum):
\begin{align}\label{eq:glob_pot}
V(\phi,&\theta,\pi_+,\pi_-)=
\frac{2{g^\prime}^2\pi_+^2\pi_-^2}{\gamma^2}\left(\frac{\Lambda_1}{\phi}\right)^
{2\gamma-2}+\left(\frac{\Lambda_1}{\phi}\right)^{2\gamma-2}\left(\frac{\phi^{
2\alpha-2}}{\Lambda_0^{2\alpha-6}}+\frac{\phi^{2\gamma-2}}{\Lambda_2^{2\gamma-6}
} \right)\nonumber\\&
+\frac{2}{\gamma}\left(\frac{\Lambda_1}{\phi}\right)^{2\gamma-2}\frac{\phi^{
\alpha+\gamma-2}}{\Lambda_0^{\alpha-3}\Lambda_2^{\gamma-3}}\cos[
(\alpha-\gamma)\theta]\nonumber\\&+\frac{g^\prime}{\sqrt{2}\gamma\Lambda_0^{
\alpha-3}}\phi^{\alpha-1}\left(\frac{\Lambda_1}{\phi}\right)^{2\gamma-2}\sqrt{
\pi_+^2\pi_-^2-\frac{8\phi^{2\gamma-2}}{{g^\prime}^2\Lambda_2^{2\gamma-6}}\sin^2
[ (\alpha-\gamma)\theta]}\nonumber\\&+\frac{g^\prime}{\sqrt{2}\gamma\Lambda_2^{
\gamma-3}}\phi^{\gamma-1}\left(\frac{\Lambda_1}{\phi}\right)^{2\gamma-2}\sqrt{
\pi_+^2\pi_-^2-\frac{8\phi^{2\alpha-2}}{{g^\prime}^2\Lambda_0^{2\alpha-6}}\sin^2
[(\alpha-\gamma)\theta]}.
\end{align}At first glance, one may worry about the square roots, however it is important to note that the above expression is only true when $\psi$ is fixed to its minimising value. Furthermore, if one examines equation (\ref{eq:sinsin}) then it is evident that as $\pi_+,\pi_-\rightarrow0$ the second term in the square root has exactly the same behaviour and so there is never a region in configuration space where the argument is negative. Minimising (\ref{eq:glob_pot}) together with the potential coming from the D-term and the $\pi_\pm$ terms in the superpotential,
\begin{equation}
 V_D+\left\vert\frac{\dd W}{\dd \pi_+}\right\vert+\left\vert\frac{\dd W}{\dd \pi_-}\right\vert= \frac{1}{2}\left(q\pi_+^2+-q\pi_-^2-\xi^2\right)^2+{g^{\prime}}^2\phi^2\left(\pi_+^2+\pi_-^2\right),
\end{equation}
with respect to $\pi_-$ one indeed finds that $\langle\pi_-\rangle=0$ is a solution. If one expands the global potential around this minimum by setting $\pi_-\rightarrow\langle\pi_-\rangle+\delta\pi_-$ then the coefficient of the $\delta\pi_-^2$ term is
\begin{equation}
 q\left(\xi^2-q\pi_+^2\right)+{g^\prime}^2|\phi|^2.
\end{equation}
In theory, this can be negative, however we have not yet finished minimising the
potential. In section \ref{sec:cosgen} we learnt that there are two possible
solutions for $\pi_+$ given by equation (\ref{eq:pi+}) when $\pi_-=0$ and so we
should check that these are indeed stable minima of the global potential. The
case where $\pi_+=0$ is clearly a minimum since the negative term vanishes. The
second case gives the coefficient as $2{g^\prime}^2|\phi|^2$ and so in either
case the coefficient is positive and the stationary point is a stable minimum.
When $\langle\pi_-\rangle=0$ equation (\ref{eq:sinsin}) gives
$\sin[(\alpha-\gamma)\theta]=0$ and hence $\cos[(\alpha-\gamma)\theta]=-1$.
Making this substitution in (\ref{eq:glob_pot}) yields the far simpler form of
the potential given in (\ref{eq:Fpot}).

\section{D-term Corrections to the Effective Potential}

In this section we briefly show how one can explore the regions of the low-energy parameter space where the corrections to the effective potential coming from the $U(1)$ symmetry breaking solution (equation (\ref{eq:dcorr})) may render the model not viable. In particular, we will explain how the sub-regions where $\phi$ passes $\Delta$ and the cosmological constant is still generated can be discerned. The corrections to the effective potential are of the form
\begin{equation}\label{eq:dcorr2}
 V_{\rm corr}=\frac{{g^\prime}^2\xi^2}{q}\phi^2-\frac{{g^\prime}^4}{2q^2}\phi^4
\end{equation}
and we shall make use of the definition \ref{eq:Gdef} and for brevity define $Z\equiv(1+z_\infty)^3$.

\subsection{Late Time Importance of the Corrections}

We can estimate the density at which each correction becomes important and we can no longer neglect them by equating each one with the magnitude of the density dependent term in turn. In this case, one finds that the field values $\phi_i$ at which the order $i$ corrections are important are
\begin{align}
\phi_2&=\left(\frac{G^2\xi^2}{x\rho}\right)^{\frac{1}{\delta-2}}\pmi^{\frac{\delta}{\delta-4}}\quad\textrm{and}\\
 \phi_4&=\left(\frac{G^4}{x\rc}\right)^{\frac{1}{\delta-4}}\pmi^{\frac{\delta}{\delta-4}}
\end{align}
respectively. Now we can always begin the cosmic evolution far enough in the past such that the density is large enough that the corrections are negligible, in which case the field evolves according to the background cosmology detailed in section \ref{sec:backcos}. As the field evolves, the coefficient of the density dependent term becomes smaller and the corrections will eventually become important. If this occurs before the field rolls past $\Delta$ then we must correct the dynamics appropriately. If, on the other hand, this occurs after the field has passed $\Delta$ then these corrections will no longer be present and we can neglect them completely. Using equations (\ref{eq:vphimin}) and (\ref{eq:vphi}), we can estimate the densities $\rho_i$ at which $\phi=\phi_i$ and the density $\rho_\Delta$ at which $\phi=\Delta$:
\begin{align}\label{eq:rhoi}\begin{split}
\rho_2^{\frac{n+2}{n+\delta}}&=\frac{M^2G^2\xi^2}{x}\left(\frac{M}{\Lambda}\right)^{\frac{4}{n}}\rho_\infty^{-\frac{\delta-2}{n+\delta}}\\\rho_4^{\frac{n+4}{n+\delta}}&=\frac{M^4G^4}{2x}\left(\frac{M}{\Lambda}\right)^{\frac{16}{n}}\rho_\infty^{-\frac{\delta-4}{n+4}}\\
 \rho_\Delta^{\frac{1}{n+\delta}}&=\frac{GM}{\xi}\left(\frac{M}{\Lambda}\right)^{\frac{4}{n}}\rho_\infty^{\frac{1}{n+\delta}}\end{split}
\end{align}
The condition that the corrections can be neglected is then $\rho_\Delta\gg\rho_i$. In this analysis we shall take ``much greater than'' to mean an order of magnitude i.e. $\rho_\Delta\ge 10\rho_i$.

\subsubsection{Mass Scales}

One must be careful that the parameters above are inter-dependent and it is important to keep track of which are fixed by specific choices of others. In the analysis of section \ref{sec:cosgen} we consider the low-energy model parameters $\{n,\delta,x,m,G,z_\infty\}$ independent, which completely fixes the derived scales $M$ and $\Lambda$ via equation \ref{eq:scales}. We start by writing equations (\ref{eq:x}) and (\ref{eq:rhoinf}) in the form
\begin{align}
 \left(\frac{\Lambda}{10^{-3}\textrm{eV}}\right)^4&=\frac{\delta x}{n}Z \label{eq:Lambda}\\ \pmi&=x^{\frac{1}{\delta}}\mu\label{eq:pmi},
\end{align}
which can be combined using equation (\ref{eq:scales}) to find
\begin{equation}\label{eq:M4n}
M^{4+n}=10^{-12}\frac{\delta}{n}Zx^{\frac{n+\delta}{\delta}}\mu^n\textrm{eV}^4.
\end{equation}
These relations can then be used to eliminate the quantities $M$ and $\Lambda$ in equation (\ref{eq:rhoi}) in favour of the low-energy parameters.

\subsection{The Quadratic Correction}

When $\rho_2\gg \rho_\Delta,\rho_4,\rc$ the effective potential is
\begin{equation}
 V_{\rm eff}(\phi)\approx\Lambda^4\left[1-\left(\frac{\phi_{\rm min}}{\phi}\right)^{\frac{n}{2}}\right]^2+G^2\xi^2\phi^2.
\end{equation}
This is minimised at field values satisfying
\begin{equation}
 \left(\frac{\pmi}{\phi}\right)^{n+2}-\left(\frac{\pmi}{\phi}\right)^{\frac{n}{2}+2}=\frac{G^2\xi^2x^{\frac{2}{\delta}}{\mu^2}}{n\Lambda^4}.
\end{equation}
When $G\xi x^{1/\delta}\mu\ll \Lambda^2$ we have $\phi\approx\pmi$ and so this case is still viable provided that $\pmi>\Delta$. If the converse is true then the minimum lies at field values
\begin{equation}
 \phi=\pmi\left(\frac{n\Lambda^4}{G^2\xi^2x^{\frac{2}{\delta}}\mu^2}\right)^{\frac{1}{n+2}}
\end{equation}
and demanding that this is larger than $\Delta$ we find that the parameters must satisfy
\begin{equation}\label{eq:quadcorr}
 \xi^{n+4}G^nx^{\frac{n}{\delta}}\mu^n>n\Lambda^4
\end{equation}
in order for the cosmological constant to be generated.

\subsection{The Quartic Correction}\label{sec:quart}

When the quartic correction is important the potential takes the following form:
\begin{equation}
 V(\phi)=\Lambda^4\left(1-\left(\frac{\phi_{\rm min}}{\phi}\right)^{\frac{n}{2}}\right)^2-\frac{G^4}{2}\phi^4.
\end{equation}
The minimum, if it exists, is given by the solution of
\begin{equation}
 \left(\frac{\phi_{\rm min}}{\phi}\right)^{\frac{n}{2}+4}-\left(\frac{\phi_{\rm min}}{\phi}\right)^{n+4}=\frac{2}{n}G^4\left(\frac{\phi_{\rm min}}{\Lambda}\right)^4
\end{equation}
and so the only possible solutions have $\phi>\phi_{\rm min}$. This means that when this correction only is important the field will always pass $\Delta$ at some time and generate a cosmological constant. With this in mind, one may wonder if the case $\pmi<\Delta$ is allowed since in this case the field can still pass $\Delta$ if the minimum lies at large enough field values. This situation is highly unnatural since once the corrections vanish the field lies at values greater than $\pmi$ and will subsequently roll backwards, reintroducing the corrections. Hubble friction will eventually reduce the amplitude of the oscillations, however this leads to a situation that is highly fine-tuned and sensitive to the initial conditions and so we exclude it.

\subsection{Simultaneous Corrections}

When both corrections are simultaneously important the effective potential, including the matter coupling, is
\begin{equation}
 V(\phi)=\Lambda^4\left(1-\left(\frac{\phi_{\rm min}}{\phi}\right)^{\frac{n}{2}}\right)^2+x\rc\left(\frac{\phi}{\phi_{\rm min}}\right)^\delta-\frac{G^4}{2}\phi^4+G^2\xi^2\phi^2.
\end{equation}
When including the density term one should technically solve the entire dynamical system in terms of $\varphi$ and $\rc(t)$, however we can glean all the information we need if we simply set $\rc=1$ eV$^4$. As mentioned above, we require that the field rolls past $\Delta$ before $\rc=1$ eV$^4$ so that the cosmological constant is generated before last scattering and this will be the case if the minimum is located at field values greater than this by last scattering. Technically, this condition is not sufficient since it only guarantees that the minimum is located at values greater than $\Delta$ by last scattering, not that the field passes $\Delta$ by this time, however the large mass of the field ensures that this approximation is sensible. Far enough in the past the corrections are unimportant and field tracks its minimum owing to this large mass. Eventually we reach the epoch where all three terms become important and pass to the regime where the density dependent term is negligible. This transition is smooth and so given the large mass we expect that the field should simply remain fixed at the new, density-independent minimum and therefore the dynamics should not differ largely from the static analysis we will employ here.

The minimum satisfies the equation
\begin{equation}
 \frac{n\Lambda^4}{\phi_{\rm min}^2}\left[\left(\frac{\phi_{\rm min}}{\phi}\right)^{\frac{n}{2}+2}-\left(\frac{\phi_{\rm min}}{\phi}\right)^{n+2}\right]+\frac{x\delta \rho_{\rm c}}{\phi_{\rm min}^2}\left(\frac{\phi_{\rm min}}{\phi}\right)^{2-\delta}+2G^2\xi^2-2G^4\phi^2=0,
\end{equation}
which must be solved numerically for the minimum given a specific set of parameters and for the same reasons given in section \ref{sec:quart}, we will impose $\pmi>\Delta$. If this has no solutions then the potential is runaway near $\rc=1$ eV$^4$ and the field will be able to pass $\Delta$. When solutions exist the parameters where the minimum occurs at field values larger than $\Delta$ are viable and those where the converse is true are not.

\bibliographystyle{jhep.bst}
\bibliography{ref}
\end{document}